\documentclass[openany]{now}
\usepackage{psfrag,verbatim,booktabs}
\usepackage{colortbl,xcolor}
\usepackage{amsmath,amsfonts}
\usepackage{natbib}

\newcommand{\BEAS}{\begin{eqnarray*}}
\newcommand{\EEAS}{\end{eqnarray*}}
\newcommand{\BEQ}{\begin{equation}}
\newcommand{\EEQ}{\end{equation}}
\newcommand{\BIT}{\begin{itemize}}
\newcommand{\EIT}{\end{itemize}}

\newcommand{\eg}{{\it e.g.}}
\newcommand{\ie}{{\it i.e.}}
\newcommand{\ones}{\mathbf 1}

\newcommand{\reals}{{\mbox{\bf R}}}

\newcommand{\Expect}{\mathbf{E}}
\newcommand{\var}{\mathop{\bf var}}
\newcommand{\diag}{\mathop{\rm diag}}

\newcommand{\cov}{\mathbf{cov}}
\newcommand{\SR}{\mathrm{SR}}
\newcommand{\IR}{\mathrm{IR}}
\newcommand{\trcost}{\phi^\mathrm{trade}}
\newcommand{\hldcost}{\phi^\mathrm{hold}}
\newcommand{\trcosthat}{\hat \phi^\mathrm{trade}}
\newcommand{\hldcosthat}{\hat \phi^\mathrm{hold}}
\newcommand{\Rp}{R^\mathrm{p}}
\newcommand{\Rphat}{\hat R^\mathrm{p}}
\newcommand{\Rehat}{\hat R^\mathrm{e}}
\newcommand{\Rahat}{\hat R^\mathrm{a}}
\newcommand{\Gp}{G^\mathrm{p}}
\newcommand{\Rb}{R^\mathrm{b}}
\newcommand{\Rep}{R^\mathrm{e}}
\newcommand{\Rap}{R^\mathrm{a}}

\newcommand{\wb}{w^\mathrm{b}}

\renewcommand{\S}{section~}

\begin{document}

\title{Multi-Period Trading via Convex Optimization}

\author{
Stephen Boyd \\
Stanford University \\
boyd@stanford.edu
\and
Enzo Busseti \\
Stanford University \\
ebusseti@stanford.edu
\and
Steven Diamond \\
Stanford University \\
stevend2@stanford.edu
\and
Ronald N. Kahn \\
Blackrock \\
ron.kahn@blackrock.com
\and
Kwangmoo Koh \\
Blackrock \\
kwangmoo.koh@blackrock.com
\and
Peter Nystrup \\
Technical University of Denmark \\
pnys@dtu.dk
\and
Jan Speth \\
Blackrock \\
jan.speth@blackrock.com
}

\frontmatter

\maketitle

\tableofcontents

\mainmatter

\begin{abstract}
We consider a basic model of multi-period trading, which can
be used to evaluate the performance of a trading strategy.
We describe a framework for single-period optimization,
where the trades in each period are found by solving a
convex optimization problem that trades off expected return,
risk, transaction cost and holding cost such as the
borrowing cost for shorting assets.
We then describe a multi-period version of the trading
method, where optimization is used to plan a sequence of
trades, with only the first one executed, using estimates
of future quantities that are unknown when the trades
are chosen.
The single-period method traces back to Markowitz;
the multi-period methods trace back to model predictive control.
\nocite{bemporad2006model,mattingley2011receding}
Our contribution is to describe the single-period and
multi-period methods in one simple framework, giving a clear
description of the development and the approximations made.
In this paper we do not address a critical component in
a trading algorithm, the predictions or forecasts
of future quantities.
The methods we describe in this paper can be thought of as
good ways to exploit predictions, no matter how they are made.
We have also developed a companion open-source software library
that implements many of the ideas and methods described in
the paper.
\end{abstract}

\chapter{Introduction}

\paragraph{Single and multi-period portfolio selection.}
Markowitz \cite{markowitz1952portfolio} was the first to formulate the choice of an
investment portfolio as an optimization problem trading off risk and return.
Traditionally, this was done independently of the cost associated
with trading, which can be significant when trades are made over
multiple periods \cite{kolm2014years}.
Goldsmith \cite{goldsmith1976transactions} was among the first to consider the
effect of transaction cost on portfolio selection in a single-period setting.
It is possible to include many other costs and constraints in a single-period
optimization formulation for portfolio selection \cite{lobo2007portfolio,moallemi2015dynamic}.

In multi-period portfolio selection, the portfolio selection problem is to choose
a sequence of trades to carry out over a set of periods.
There has been much research on this topic since the work of
Samuelson \cite{samuelson1969liftetime} and
Merton \cite{merton1969lifetime, merton1971optimum}.
Constantinides \cite{constantinides1979multiperiod} extended Samuelson's
discrete-time formulation to problems with proportional transaction costs.
Davis and Norman \cite{davis1990portfolio} and Dumas and Lucian \cite{dumas1991exact}
derived similar results for the continuous-time formulation.
Transaction costs, constraints, and time-varying forecasts are more
naturally dealt with in a multi-period setting.
Following Samuelson and Merton, the literature on multi-period portfolio selection is
predominantly based on dynamic programming, which properly takes into
account the idea of recourse and updated information available as the sequence of
trades are chosen (see \cite{garleanu2013dynamic} and references therein).
Unfortunately, actually carrying out dynamic programming
for trade selection is impractical, except for some very special or small cases,
due to the `curse of dimensionality' \cite{bellman1956dynamic,boyd2014performance}.
As a consequence, most studies include only a very limited number of assets
and simple objectives and constraints.
A large literature studies multi-period portfolio selection
in the absence of transaction cost (see, \eg, \cite{campbell2002strategic} and
references therein); in this special case, dynamic programming is tractable.

For practical implementation, various approximations of the
dynamic programming approach are often used,
such as approximate dynamic programming,
or even simpler formulations that generalize the
single-period formulations to multi-period optimization problems
\cite{boyd2014performance}.
We will focus on these simple multi-period methods in this paper.
While these simplified approaches can be criticized for only
approximating the full dynamic programming trading policy,
the performance loss is likely very small in practical problems,
for several reasons.
In \cite{boyd2014performance}, the authors developed a numerical bounding
method that quantifies the loss of optimality when using a simplified
approach, and found it to be very small in numerical examples.
But in fact, the dynamic programming formulation is
itself an approximation, based on assumptions (like independent
or identically distributed returns) that need not hold well
in practice, so the idea of an `optimal strategy' itself
should be regarded with some suspicion.

\paragraph{Why now?}
What is different now, compared to 10, 20, or 30 years ago,
is vastly more powerful computer power, better algorithms,
specification languages for optimization, and access to much more
data.
These developments have changed
how we can use optimization in multi-period investing.  In
particular, we can now quickly run full-blown optimizations, run multi-period
optimizations, and search over hyper-parameter spaces in back-tests.  We can run
end-to-end analyses, indeed many at a time in parallel.
Earlier generations of investment researchers, relying on
computers much less powerful than today, relied much more on separate models
and analyses to estimate parameter values, and tested signals using simplified
(usually unconstrained) optimizations.

\paragraph{Goal.}
In this tutorial paper we consider multi-period investment and trading.
Our goal is to describe a simple model that takes into account
the main practical issues that arise,
and several simple and practical frameworks
based on solving convex optimization problems \cite{boyd2004convex} that determine
the trades to make.
We describe the approximations made, and briefly discuss how the methods
can be used in practice.
Our methods do not give a complete trading system,
since we leave a critical part unspecified:
Forecasting future returns, volumes, volatilities, and other important
quantities (see, \eg, \cite{grinold1999active}).
This paper describes good practical methods that can be used to trade, given forecasts.

The optimization-based trading methods we describe are practical
and reliable when the problems to be solved are convex.
Real-world single-period convex problems with thousands of assets can be
solved using generic algorithms in well under a second,
which is critical for evaluating
a proposed algorithm with historical or simulated data, for many values of
the parameters in the method.

\paragraph{Outline.}
We start in chapter \ref{s-model} by describing a simple model of
multi-period trading, taking into account returns, trading costs,
holding costs, and (some) corporate actions.
This model allows us to carry out simulation, used
for what-if analyses, to see what would have happened
under different conditions, or with a different trading strategy.
The data in simulation can be realized past data (in a \emph{back-test})
or simulated data that did not occur, but could have occurred
(in a \emph{what-if simulation}), or data chosen to be particularly
challenging (in a \emph{stress-test}).
In chapter \ref{s-metrics} we review several common metrics used to evaluate
(realized or simulated)
trading performance, such as active return and risk with respect to
a benchmark.

We then turn to optimization-based trading strategies.
In chapter \ref{s-spo} we describe \emph{single-period optimization} (SPO),
a simple but effective framework for trading based on
optimizing the portfolio performance over a single period.
In chapter \ref{s-mpo} we consider \emph{multi-period optimization} (MPO),
where the trades are chosen by solving an optimization problem
that covers multiple periods in the future.

\paragraph{Contribution.}
Most of the material that appears in this paper has appeared
before, in other papers, books, or EE364A, the Stanford course on
convex optimization.  Our contribution is to collect in one place
the basic definitions, a careful description of the model,
and discussion of how convex optimization
can be used in multi-period trading, all in a
common notation and framework.  Our goal is not to survey all the work
done in this and related areas, but rather to give a unified,
self-contained treatment.  Our focus is not on theoretical issues,
but on practical ones that arise in multi-period trading.
To further this goal,
we have developed an accompanying open-source software library
implemented in Python, and available at
\begin{quote}
\url{https://github.com/cvxgrp/cvxportfolio}.
\end{quote}

\paragraph{Target audience.}
We assume that the reader has a background in the basic ideas of
quantitative portfolio selection, trading, and finance, as described for
example in the books by
Grinold \& Kahn \cite{grinold1999active}, Meucci \cite{meucci2005risk}, or Narang \cite{narang2013inside}.
We also assume that the reader has seen some basic
mathematical optimization, specifically convex optimization \cite{boyd2004convex}.
The reader certainly does not need to know more than the very basic ideas
of convex optimization, for example the overview material covered in chapter 1 of
this book (\cite{boyd2004convex}).
In a nutshell, our target reader is a quantitative trader,
or someone who works with or for, or employs, one.

\chapter{The Model}\label{s-model}
In this section we set the notation and give some detail of
our simplified model of multi-period trading.
We develop our basic dynamic model of trading, which tells us
how a portfolio and associated cash account change over time,
due to trading, investment gains, and various costs associated with
trading and holding portfolios.
The model developed in this section
is independent of any method for choosing or evaluating the
trades or portfolio strategy, and independent of any method used to
evaluate the performance of the trading.

\section{Portfolio asset and cash holdings}
\paragraph{Portfolio.}
We consider a portfolio of holdings in $n$ assets,
plus a cash account, over a finite time horizon, which is
divided into discrete time periods labeled $t=1,\ldots,T$.
These time periods need not be
uniformly spaced in real time or be of equal length; for example when they
represent trading days, the periods are one (calendar) day during
the week and three (calendar) days over a weekend.
We use the label $t$ to refer to both
a point in time, the beginning of time period $t$,
as well as the time interval from time $t$ to $t+1$.
The time period in our model is arbitrary, and could be daily, weekly,
or one hour intervals, for example.  We will occasionally give examples
where the time indexes trading days, but the same notation and model
apply to any other time period.

Our investments will be in a \emph{universe} of $n$ assets,
along with an associated cash account.
We let $h_t \in \reals^{n+1}$ denote the portfolio
(or vector of positions or holdings) at the beginning of time period $t$,
where $(h_t)_i$ is the \emph{dollar
value} of asset $i$ at the beginning of time period $t$, with $(h_t)_i <0$
meaning a short position in asset $i$, for $i=1, \ldots, n$.
The portfolio is \emph{long-only} when the asset holdings are all
nonnegative, \ie, $(h_t)_i \geq 0$ for $i=1, \ldots, n$.

The value of $(h_t)_{n+1}$ is the \emph{cash balance},
with $(h_t)_{n+1}<0$ meaning that money is owed (or borrowed).
The dollar value for the assets
is determined using the reference prices $p_t \in \reals^n_+$,
defined as the average of the bid and ask prices at the beginning of
time period $t$.
When $(h_t)_{n+1}=0$, the portfolio is \emph{fully invested}, meaning that
we hold (or owe) zero cash, and all our holdings (long and short) are in
assets.

\paragraph{Total value, exposure, and leverage.}
The total value (or \emph{net asset value}, NAV)
$v_t$ of the portfolio, in dollars,
at time $t$ is $v_t = \ones^Th_t$, where
$\ones$ is the vector with all entries one.
(This is not quite the amount of cash the portfolio would yield on liquidation,
due to transaction costs, discussed below.)
Throughout this paper we will assume that $v_t>0$, \ie, the total
portfolio value is positive.

The vector
\[
(h_t)_{1:n} = ((h_t)_1, \ldots, (h_t)_n)
\]
gives the asset holdings.
The \emph{gross exposure} can be expressed as
\[
\|(h_t)_{1:n}\|_1 = |(h_t)_1| + \cdots + |(h_t)_n|,
\]
the sum of the absolute values of the asset positions.
The \emph{leverage} of the portfolio is the gross exposure divided by
the value, $\|(h_t)_{1:n}\|_1/v_t$.
(Several other definitions of leverage are also used, such as
half the quantity above.)
The leverage of a fully invested long-only portfolio is one.

\paragraph{Weights.}
We will also describe the portfolio using weights or fractions
of total value.
The \emph{weights} (or weight vector) $w_t \in \reals^{n+1}$
associated with the portfolio $h_t$ are defined as
$w_t = h_t/v_t$. (Recall our assumption that $v_t>0$.)
By definition the weights sum to one, $\ones^Tw_t=1$, and are unitless.
The weight $(w_t)_{n+1}$ is the fraction of the total portfolio
value held in cash.
The weights are all nonnegative when the asset positions
are long and the cash balance is nonnegative.
The dollar value holdings vector can be expressed in terms of the weights
as $h_t = v_t w_t$.
The leverage of the portfolio can be expressed in terms of the
weights as $\|w_{1:n}\|_1$, the $\ell_1$-norm of the asset weights.

\section{Trades}
\paragraph{Trade vector.}
In our simplified model we assume that all trading, \ie,
buying and selling of assets, occurs at the beginning of each time period.
(In reality the trades would likely be spread over at least some part of the period.)
We let $u_t \in \reals^n$ denote the dollar values of the
trades, at the current price: $(u_t)_i>0$ means we buy
asset $i$ and $(u_t)_i<0$ means we sell
asset $i$, at the beginning of time period $t$, for $i=1, \ldots, n$.
The number $(u_t)_{n+1}$ is the amount we put into the cash account (or take out,
if it is negative).
The vector $z_t = u_t/v_t$ gives the trades normalized by the total value.
Like the weight vector $w_t$, it is unitless.

\paragraph{Post-trade portfolio.}
The post-trade portfolio is denoted
\[
h_t^+ = h_t + u_t, \quad t=1,\ldots, T.
\]
This is the portfolio in time period $t$ immediately after trading.
The post-trade portfolio value is $v_t^+ = \ones^T h_t^+$.
The change in total portfolio value from the trades is given by
\[
v_t ^+ - v_t = \ones^T h_t^+ - \ones^T h_t = \ones^T u_t.
\]
The vector $(u_t)_{1:n}\in \reals^n$ is the set of (non-cash) asset trades.
Half its $\ell_1$-norm $\|(u_t)_{1:n}\|_1/2$ is the \emph{turnover}
(in dollars) in period $t$.
This is often expressed as a percentage of total value, as
$\|(u_t)_{1:n}\|_1/(2v_t) = \|z_{1:n}\|_1/2$.

We can express the post-trade portfolio, normalized by the portfolio
value, in terms of the weights $w_t=h_t/v_t$ and normalized trades as
\BEQ\label{e-wt+}
h_t^+/v_t = w_t + z_t.
\EEQ
Note that this normalized quantity does not necessarily add up to one.

\section{Transaction cost}
The trading incurs a trading or transaction cost (in dollars), which we denote as $\trcost_t(u_t)$,
where $\trcost_t: \reals^{n+1} \to \reals$ is the (dollar) transaction cost function.
We will assume that $\trcost_t$ does not depend on $(u_t)_{n+1}$, \ie,
there is no transaction cost associated with the cash account.
To emphasize this we will sometimes write the transaction cost as
$\trcost_t((u_t)_{1:n})$.
We assume that $\trcost_t(0)=0$, \ie,
there is no transaction cost when we do not trade.
While $\trcost_t(u_t)$ is typically nonnegative, it can be negative in some cases,
discussed below.
We assume that the transaction cost function $\trcost_t$ is \emph{separable},
which means it has the form
\[
\trcost_t(x) = \sum_{i=1}^n (\trcost_t)_i(x_i),
\]
\ie, the transaction cost breaks into a sum of transaction costs
associated with the individual assets.
We refer to $(\trcost_t)_i$, which is a function from $\reals$ into $\reals$,
as the transaction cost function for asset $i$, period $t$.
We note that some authors have used models of transaction cost which are
not separable, for example Grinold's quadratic dynamic
model \cite{grinold2006dynamic}.

\paragraph{A generic transaction cost model.}
A reasonable model for the scalar transaction cost functions $(\trcost_t)_i$ is
\BEQ\label{e-tcost}
x \mapsto a|x| + b \sigma \frac{|x|^{3/2}}{V^{1/2}} + cx,
\EEQ
where $a$, $b$, $\sigma$, $V$, and $c$ are real numbers described below,
and $x$ is a dollar trade amount \cite{grinold1999active}.
The number $a$ is one half the bid-ask spread for the asset
at the beginning of the time period, expressed as a fraction of the
asset price (and so is unitless).
We can also include in this term
broker commissions or fees which are a linear
function of the number of shares (or dollar value) bought or sold.
The number $b$ is a positive constant with unit inverse dollars.
The number $V$ is the total market volume traded
for the asset in the time period, expressed in dollar value,
so $|x|^{3/2}/V^{1/2}$ has units of dollars.
The number $\sigma$ the corresponding price volatility
(standard deviation) over recent time periods, in dollars.
According to a standard rule of thumb,
trading one day's volume moves the
price by about one day's volatility, which suggests that the value of
the number $b$ is around one.
(In practice, however, the value of $b$ is determined
by fitting the model above to data on realized transaction costs.)
The number $c$ is used to create asymmetry in the transaction
cost function.  When $c=0$, the transaction cost is the same for
buying and selling; it is a function of $|x|$.  When $c>0$,
it is cheaper to sell than to buy the asset,
which generally occurs in a market where
the buyers are providing more liquidity than the sellers
(\eg, if the book is not balanced in a limit order exchange).
The asymmetry in transaction cost can also be used to model
price movement during trade execution.
Negative transaction cost can occur when $|c| > |a|$.
The constants in the transaction cost model \eqref{e-tcost} vary with asset,
and with trading period, \ie, they are indexed by $i$ and $t$.

\paragraph{Normalized transaction cost.}
The transaction cost model \eqref{e-tcost} is in dollars.
We can normalize it by $v_t$, the total portfolio value, and
express it in terms of $z_i$, the normalized trade of asset $i$,
resulting in the function (with $t$ suppressed for simplicity)
\BEQ\label{e-tcost-weights}
a_i|z_i| + b_i \sigma_i \frac{|z_i|^{3/2}}{(V_i/v)^{1/2}} + c_iz_i.
\EEQ
The only difference with \eqref{e-tcost} is that we use
the normalized asset volume $V_i/v$ instead of the dollar volume $V_i$.
This shows that the same transaction cost formula can be used to express
the dollar transaction cost as a function of the dollar trade, with
the volume denoted in dollars,
or the normalized transaction cost as a function of the normalized
trade, with the volume normalized by the portfolio value.

With some abuse of notation, we will write the normalized transaction cost
in period $t$ as $\trcost_t(z_t)$.  When the argument to the transaction cost
function is normalized, we use the version where asset volume is also normalized.
The normalized transaction cost $\trcost_t(z_t)$ depends on the portfolio value
$v_t$, as well as the current values of the other parameters,
but we suppress this dependence to lighten the notation.

\paragraph{Other transaction cost models.}
Other transaction cost models can be used.  Common variants include
a piecewise linear model, or adding a term that is quadratic in the trade value $z_i$
\cite{almgren2001optimal,grinold2006dynamic,garleanu2013dynamic}.
Almost all of these are convex functions. (We discuss this later.)
An example of a transaction cost term that is not convex is
a fixed fee for any nonzero trading in an asset.
For simulation, however, the transaction cost function can be arbitrary.

\section{Holding cost}
We will hold the post-trade portfolio $h_t^+$ over the $t$th period.
This will incur a holding-based cost (in dollars) $\hldcost_t(h_t^+)$,
where $\hldcost_t:\reals^{n+1} \to \reals$ is the holding cost function.
Like transaction cost, it is typically nonnegative, but it can also be
negative in certain cases, discussed below.
The holding cost can include a factor related to the length of the period;
for example if our periods are trading days, but holding costs are assessed on
all days (including weekend and holidays), the friday holding cost might be
multiplied by three.
For simplicity, we will assume that the holding cost function
does not depend on the post-trade cash balance $(h_t^+)_{n+1}$.

A basic holding cost model includes
a charge for borrowing assets when going short, which has the form
\BEQ\label{e-hld-cost}
\hldcost_t(h_t^+) =  s_t^T (h^+_t)_-,
\EEQ
where $(s_t)_i \geq 0$ is the borrowing fee,
in period $t$, for shorting asset $i$,
and $(z)_- =\max\{-z,0\}$ denotes the negative part of a number $z$.
This is the fee for shorting
the post-trade assets, over the investment period, and here we are paying this
fee in advance, at the beginning of the period.
Our assumption that the holding cost does not depend on the cash account
requires $(s_t)_{n+1} =0$.  But we can include a cash borrow cost if needed,
in which case $(s_t)_{n+1}>0$.  This is the premium for borrowing, and not
the interest rate, which is included in another part of our model,
discussed below.

The holding cost \eqref{e-hld-cost}, normalized by portfolio value,
can be expressed in terms of weights and normalized trades as
\BEQ\label{e-hld-cost-weights}
\hldcost(h_t^+)/v_t =  s_t^T (w_t + z_t)_-.
\EEQ
As with the transaction cost, with some abuse of notation
we use the same function symbol to denote the normalized holding cost,
writing the quantity above as $\hldcost_t(w_t+z_t)$.
(For the particular form of holding cost described above, there is no abuse of
notation since $\hldcost_t$ is the same when expressed in dollars or
normalized form.)

More complex holding cost functions
arise, for example when the assets include ETFs (exchange traded funds).
A long position incurs a fee proportional to $h_i$;
when we hold a short position, we \emph{earn} the same fee.  This is readily modeled
as a linear term in the holding cost.  (We can in addition have a standard fee
for shorting.)  This leads to a holding cost of the form
\[
\hldcost_t (w_t+z_t) = s_t^T (w_t + z_t)_- + f_t^T(w_t+z_t),
\]
where $f_t$ is a vector with $(f_t)_i$ representing the per-period
management fee for asset $i$, when asset $i$ is an ETF.

Even more complex holding cost models can be used.
One example is a piecewise linear
model for the borrowing cost, which increases the marginal borrow charge rate
when the short position exceeds some threshold.
These more general holding cost functions are almost always convex.
For simulation, however, the holding cost function can be arbitrary.

\section{Self-financing condition}
We assume that no external cash is put into or taken out of the portfolio,
and that the trading and holding costs are paid from the cash account
at the beginning of the period.
This \emph{self-financing condition} can be expressed as
\BEQ\label{e-self-financing}
\ones^Tu_t + \trcost_t(u_t) + \hldcost_t(h^+_t) = 0.
\EEQ
Here $-\ones^Tu_t$ is the total cash out of the portfolio from the trades;
\eqref{e-self-financing} says that this cash out must balance the cash cost
incurred, \ie, the transaction cost plus the holding cost.
The self-financing condition implies
$v_t^+ = v_t - \trcost_t(u_t) - \hldcost_t(h^+_t)$, \ie,
the post-trade value is the pre-trade value minus the transaction
and holding costs.

The self-financing condition \eqref{e-self-financing} connects the cash trade
amount $(u_t)_{n+1}$ to the asset trades, $(u_t)_{1:n}$, by
\BEQ\label{e-cash-trade}
(u_t)_{n+1} = - \left( \ones^T(u_t)_{1:n} + \trcost_t((h_t+u_t)_{1:n}) +
\hldcost_t((u_t)_{1:n}) \right).
\EEQ
Here we use the assumption that the transaction and holding costs do not depend
on the $n+1$ (cash) component by explicitly writing the argument as the first $n$ components,
\ie, those associated with the (non-cash) assets.
The formula \eqref{e-cash-trade} shows that if we are given the trade values for the
non-cash assets, \ie, $(u_t)_{1:n}$, we can find the cash trade value
$(u_t)_{n+1}$
that satisfies the self-financing condition \eqref{e-self-financing}.

We mention here a subtlety that will come up later.
A trading algorithm chooses the asset trades $(u_t)_{1:n}$
before the transaction cost function $\trcost_t$ and (possibly) the
holding cost function $\hldcost_t$ are known.
The trading algorithm must use \emph{estimates} of these
functions to make its choice of trades.
The formula \eqref{e-cash-trade} gives the cash trade amount that is
\emph{realized}.

\paragraph{Normalized self-financing.}
By dividing the dollar self-financing condition \eqref{e-self-financing}
by the portfolio value $v_t$,
we can express the self-financing condition in terms of weights and
normalized trades as
\[
\ones^Tz_t + \trcost_t(v_t z_t)/v_t + \hldcost_t(v_t(w_t + z_t))/v_t = 0,
\]
where we use $u_t = v_t z_t$ and $h_t^+=v_t (w_t+z_t)$, and the
cost functions above are the dollar value versions.
Expressing the costs in terms of normalized values we get
\BEQ\label{e-self-financing-weights}
\ones^Tz_t + \trcost_t(z_t) + \hldcost_t(w_t + z_t) = 0,
\EEQ
where here the costs are the normalized versions.

As in the dollar version, and assuming that the costs do not depend on the cash
values, we can express the cash trade value $(z_t)_{n+1}$ in terms of the
non-cash asset trade values $(z_t)_{1:n}$ as
\BEQ\label{e-cash-trade-weights}
(z_t)_{n+1} = - \left( \ones^T(z_t)_{1:n} + \trcost_t((w_t+ z_t)_{1:n}) +
\hldcost_t((z_t)_{1:n}) \right).
\EEQ

\section{Investment}
The post-trade portfolio and cash are invested
for one period, until the beginning of the next time period.
The portfolio at the next time period is given by
\[
h_{t+1} = h_t^+ + r_t \circ h_t^+ = (\ones+r_t) \circ
h_t^+, \quad t=1,\ldots, T-1,
\]
where $r_t \in \reals^{n+1}$ is the vector of
asset and cash returns from period $t$ to period $t+1$ and $\circ$ denotes
Hadamard (elementwise) multiplication of vectors.
The return of asset $i$ over period $t$ is defined as
\[
(r_t)_i=  \frac{(p_{t+1})_i-(p_t)_i}{(p_t)_i}, \quad i=1, \ldots, n,
\]
the fractional increase in the asset price over the investment period.
We assume here that the prices and returns are adjusted to include the
effects of stock splits and dividends.
We will assume that the prices are nonnegative, so $\ones+r_t \geq 0$
(where the inequality means elementwise).
We mention an alternative to our definition above, the \emph{log-return},
\[
\log \frac{(p_{t+1})_i}{(p_t)_i} =
\log (1+(r_t)_i), \quad i=1, \ldots, n.
\]
For returns that are small compared to one, the log-return is
very close to the return defined above.

The number $(r_t)_{n+1}$ is the return to cash, \ie, the risk-free interest rate.
In the simple model, the cash interest rate is the same for cash deposits and
loans.  We can also include a premium for borrowing cash (say) in the
holding cost function, by taking $(s_t)_{n+1}>0$ in \eqref{e-hld-cost}.
When the asset trades $(u_t)_{1:n}$ are chosen, the asset returns $(r_t)_{1:n}$
are not known.
It is reasonable to assume that the cash interest rate $(r_t)_{n+1}$ is known.

\paragraph{Next period portfolio value.}
For future reference we work out some useful formulas for the next period
portfolio value.
We have
\BEAS
v_{t+1} &=& \ones^T h_{t+1}\\
&=& (\ones + r_t)^T h_t^+\\
&=& v_t + r_t^T h_t + (\ones+r_t)^Tu_t\\
&=& v_t + r_t^T h_t + r_t^Tu_t - \trcost_t(u_t) - \hldcost_t(h_t^+).
\EEAS

\paragraph{Portfolio return.}
The \emph{portfolio realized return} in period $t$ is defined as
\[
\Rp_t = \frac{v_{t+1} - v_t}{v_t},
\]
the fractional increase in portfolio value over the period.
It can be expressed as
\BEQ
\label{e-return}
\Rp_t =
r_t^T w_t +r _t^T z_t -
\trcost_t(z_t) - \hldcost_t(w_t + z_t).
\EEQ
This is easily interpreted. The portfolio return
over period $t$ consists of four parts:
\BIT
\item $r_t^Tw_t$ is the portfolio return
without trades or holding cost,
\item $r_t^Tz_t$ is the return on the trades,
\item $-\trcost_t(z_t)$ is the transaction cost,
and \item $-\hldcost_t(w_t+z_t)$ is the holding cost.
\EIT

\paragraph{Next period weights.}
We can derive a formula for the next period weights $w_{t+1}$ in
terms of the current weights $w_t$ and the normalized trades $z_t$, and the return
$r_t$, using the equations above.  Simple algebra gives
\BEQ\label{e-wt1}
w_{t+1} = \frac{1}{1+\Rp_t} (\ones + r_t) \circ (w_t+z_t).
\EEQ
By definition, we have $\ones^Tw_{t+1}=1$.
This complicated formula reduces to $w_{t+1}=w_t+z_t$ when $r_t=0$.
We note for future use that when the per-period returns are
small compared to one, we have $w_{t+1}\approx w_t +z_t$.

\section{Aspects not modeled}

We list here some aspects of real trading that our model ignores, and discuss
some approaches to handle them if needed.

\paragraph{External cash.}  Our self-financing condition \eqref{e-self-financing}
assumes that no external cash enters or leaves the portfolio.  We can easily
include external deposits and withdrawals of cash by replacing
the right-hand side of \eqref{e-self-financing} with the external cash
put into the account (which is positive for cash deposited into the account
and negative for cash withdrawn).

\paragraph{Dividends.}  Dividends are usually included in the asset
return, which implicitly means they are re-invested.
Alternatively we can include cash dividends from assets in the
holding cost, by adding
the term $-d_t^T h_t$, where $d_t$ is the vector of dividend rates
(in dollars per dollar of the asset held) in period $t$.
In other words, we can treat cash dividends as negative holding costs.

\paragraph{Non-instant trading.} Our model assumes all trades are
carried out instantly at the beginning of each investment period,
but the trades are really executed over some fraction of the period.
This can be modeled using the linear term in the transaction
cost, which can account for the movement of the price during the
execution.
We can also change the dynamics equation
\[
h_{t+1} = (\ones + r_t) \circ (h_t + u_t)
\]
to
\[
h_{t+1} = (\ones + r_t) \circ h_t + (1-\theta_t/2) (\ones + r_t)\circ u_t,
\]
where $\theta_t$ is the fraction of the period over which the trades occur.
In this modification, we do not get the full period return on the
trades when $\theta_t>0$, since we are moving into the position as the
price moves.

The simplest method to handle non-instant trading is to use a shorter
period.  For example if we are interested in daily trading, but the trades are
carried out over the whole trading day and we wish to model this effect,
we can move to an hourly model.

\paragraph{Imperfect execution.}  Here we distinguish between $u^\mathrm{req}_t$,
the requested trade, and $u_t$, the actual realized trade.
In a back-test simulation we might assume that some
(very small) fraction of the requested trades are only partially completed.

\paragraph{Multi-period price impact.}
This is the effect of a
large order in one period affecting the asset price in future periods.
In our model the transaction cost is only a function of the current
period trade vector, not previous ones.

\paragraph{Trade settlement.}
In trade settlement we keep track of cash from trades one day and two days
ago (in daily simulation),
as well as the usual (unencumbered) cash account
which includes all cash from trades that occurred three or more
days ago, which have already settled.
Shorting expenses come from the unencumbered cash, and trade-related
cash moves immediately into the one day ago category (for daily trading).



\paragraph{Merger/acquisition.}
In a certain period one company buys another, converting the
shares of the acquired
company into shares of the acquiring company at some rate.
This modifies the asset holdings update.  In a cash buyout, positions in the
acquired company are converted to cash.

\paragraph{Bankruptcy or dissolution.}
The holdings in an asset are reduced to zero, possibly with a cash payout.

\paragraph{Trading freeze.}
A similar action is a trading freeze,
where in some time periods an asset cannot be bought, or sold, or both.

\section{Simulation}
Our model can be used to simulate the evolution of a portfolio
over the periods $t=1,\ldots, T$.
This requires the following data, when the standard model described above is used.
(If more general transaction or holding cost functions are used, any data
required for them is also needed.)
\BIT\itemsep -2pt
\item Starting portfolio and cash account values, $h_1\in \reals^{n+1}$.
\item Asset trade vectors $(u_t)_{1:n}$.
The cash trade value $(u_t)_{n+1}$ is determined from the self-financing
condition by \eqref{e-cash-trade}.
\item Transaction cost model parameters $a_t \in \reals^n$,
$b_t \in \reals^n$, $c_t\in \reals^n$,
$\sigma_t \in \reals^n$, and $V_t \in \reals^n$.
\item Shorting rates $s_t\in \reals^n$.
\item Returns $r_t\in \reals^{n+1}$.
\item Cash dividend rates $d_t\in \reals^n$,
if they are not included in the returns.
\EIT

\paragraph{Back-test.}
In a back-test the values would be past realized values,
with $(u_t)_{1:n}$ the trades proposed by the trading algorithm being
tested.  Such a test estimates what the evolution of the portfolio
would have been with different trades or a different trading algorithm.
The simulation determines the portfolio and cash account values
over the simulation period, from which other metrics, described in chapter
\ref{s-metrics}
below, can be computed.
As a simple example, we can compare the performance
of re-balancing to a given target portfolio daily, weekly, or quarterly.

A simple but informative back-test is to simulate the portfolio
evolution using the actual trades that were executed in a portfolio.
We can then compare the actual and simulated or predicted
portfolio holdings and total value over some time period.
The true and simulated portfolio values
will not be identical, since our model relies on
estimates of transaction and holding costs,
assumes instantaneous trade execution, and so on.

\paragraph{What-if simulations.}
In a what-if simulation, we change the data used to carry out the
simulation, \ie, returns, volumes, and so on.
The values used are ones that (presumably) could have occurred.
This can be used to stress-test a trading algorithm, by using data
that did not occur, but would have been very challenging.

\paragraph{Adding uncertainty in simulations.}
Any simulation of portfolio evolution relies on models of
transaction and holding costs, which in turn depend on parameters.
These parameters are not known exactly, and in any case,
the models are not exactly correct.
So the question arises, to what extent should we trust our simulations?
One simple way to check this is to carry out multiple simulations,
where we randomly perturb the model parameters by reasonable
amounts.  For example, we might vary the daily volumes
from their true (realized) values by 10\% each day.  If
simulation with parameters that are perturbed by reasonable amounts
yields divergent results, we know that (unfortunately) we cannot
trust the simulations.

\chapter{Metrics}
\label{s-metrics}
Several generic performance metrics can be used to evaluate the
portfolio performance.

\section{Absolute metrics}
We first consider metrics that measure the growth of portfolio value in absolute terms,
not in comparison to a benchmark portfolio or the risk-free rate.

\paragraph{Return and growth rate.}
The average realized return over periods $t=1,\ldots, T$ is
\[
\overline \Rp =  \frac{1}{T} \sum_{t=1}^{T} \Rp_t.
\]
An alternative measure of return is the \emph{growth rate} (or log-return)
of the portfolio in period $t$, defined as
\[
\Gp_t = \log (v_{t+1} / v_t ) = \log (1+\Rp_t).
\]
The average growth rate of the portfolio
is the average value of $\Gp_t$ over the periods $t=1, \ldots, T$.
For per-period returns that are small compared to one (which is
almost always the case in practice) $\Gp_t$ is very close to $\Rp_t$.

The return and growth rates given above are per-period. For interpretability
they are typically \emph{annualized} \cite{bacon2008practical}:
Return and growth rates are multiplied by $P$,
where $P$ is the number of periods in one year.
(For periods that are trading days, we have $P\approx 250$.)

\paragraph{Volatility and risk.}
The realized \emph{volatility} is the standard deviation of the portfolio
return time series,
\[
\sigma^\mathrm{p} = \left( \frac{1}{T} \sum_{t=1}^{T} (\Rp_t - \overline \Rp)^2
\right)^{1/2}.
\]
(This is the maximum-likelihood estimate; for an unbiased estimate we replace
$1/T$ with $1/(T-1)$).
The square of the volatility is the \emph{quadratic risk}.
When $\Rp_t$ are small (in comparison to $1$),
a good approximation of the quadratic risk is the
second moment of the return,
\[
(\sigma^\mathrm{p})^2 \approx
\frac{1}{T} \sum_{t=1}^{T} (\Rp_t)^2.
\]


The volatility and quadratic risk given above are per-period. For interpretability
they are typically annualized.  To get the annualized values
we multiply volatility by $\sqrt P$, and quadratic risk by $P$.  (This scaling
is based on the idea that the returns in different periods are
independent random variables.)

\section{Metrics relative to a benchmark}
\label{s-relativemetrics}

\paragraph{Benchmark weights.}
It is common to measure the portfolio performance against a
\emph{benchmark}, given as a set of weights
$\wb_t \in \reals^{n+1}$, which are fractions of the assets (including cash),
and satisfy $\ones^T \wb_t=1$.
We will assume the benchmark weights are nonnegative, \ie,
the entries in $\wb_t$ are nonnegative.
The benchmark weight $\wb_t = e_{n+1}$
(the unit vector with value 0 for all entries
except the last, which has value 1) represents the cash, or risk-free, benchmark.
More commonly the benchmark consists of a particular set of assets
with weights proportional to their capitalization.
The benchmark return in period $t$ is $\Rb_t = r_t^T\wb_t$.
(When the benchmark is cash, this is the risk-free interest rate $(r_t)_{n+1}$.)

\paragraph{Active and excess return.}
The \emph{active} return (of the portfolio, with respect to a benchmark) is given by
\[
\Rap_t = \Rp_t - \Rb_t.
\]
In the special case when the benchmark consists of cash (so that the benchmark
return is the risk-free rate) this is known as \emph{excess return}, denoted
\[
\Rep_t = \Rp_t - (r_t)_{n+1}.
\]
We define the average \emph{active return} $\overline \Rap$, relative to the benchmark,
as the average of $\Rap_t$.
We have
\BEAS
\Rap_t &=& \Rp_t - \Rb_t \\
&=& r_t^T \left( w_t - \wb_t \right) + r_t^T z_t
-\trcost_t(z_t) - \hldcost_t(w_t + z_t).
\EEAS
Note that if $z_t=0$ and $w_t = \wb_t$, \ie, we hold the benchmark weights
and do not trade, the active return is zero.
(This relies on the assumption that the benchmark weights are nonnegative,
so $\hldcost_t(\wb_t)=0$.)

\paragraph{Active risk.}
The standard deviation of $\Rap_t$, denoted $\sigma^\mathrm{a}$,
is the risk relative to the benchmark, or \emph{active risk}.
When the benchmark is cash, this is the excess risk $\sigma^\mathrm{e}$.
When the risk-free interest rate is constant, this is the
same as the risk $\sigma^\mathrm{p}$.

\paragraph{Information and Sharpe ratio.}
The (realized) \emph{information ratio} (IR) of the portfolio relative
to a benchmark is the average of the
active returns $\overline \Rap$ over the standard deviation of the
active returns $\sigma^\mathrm{a}$ \cite{grinold1999active},
\[
\IR = \overline \Rap / \sigma^\mathrm{a}.
\]
In the special case of a cash benchmark this is known as \emph{Sharpe ratio} (SR)
\cite{sharpe1966,sharpe1994}
\[
\SR = \overline \Rep / \sigma^\mathrm{e}.
\]
Both $\IR$ and $\SR$ are typically given using the annualized
values of the return and risk \cite{bacon2008practical}.

\chapter{Single-Period Optimization}
\label{s-spo}

In this section we consider optimization-based trading strategies
where at the beginning of period $t$, using all the data available,
we determine the asset portion of the current trade vector $(u_t)_{1:n}$
(or the normalized asset trades $(z_t)_{1:n}$).
The cash component of the trade vector $(z_t)_{n+1}$
is then determined by the self-financing equation \eqref{e-cash-trade-weights},
once we know the realized costs.
We formulate this as a convex optimization problem,
which takes into account the portfolio performance over one period,
the constraints on the portfolio, and investment risk (described below).
The idea goes back to Markowitz \cite{markowitz1952portfolio},
who was the first to formulate the choice of a portfolio
as an optimization problem.
(We will consider multi-period optimization in the next section.)


When we choose $(z_t)_{1:n}$, we do not know $r_t$ and the other
market parameters (and therefore the transaction cost function $\trcost_t$), so
instead we must rely on estimates of these quantities and functions.
We will denote an estimate of the quantity or function $Z$, made at the beginning
of period $t$
(\ie, when we choose $(z_t)_{1:n}$), as $\hat Z$.
For example $\trcosthat_t$ is our
estimate of the current period transaction cost function
(which depends on the market volume and other parameters,
which are predicted or estimated).
The most important quantity that we estimate is the return over the
current period $r_t$, which we denote as $\hat r_t$.
(Return forecasts are sometimes called \emph{signals}.)
If we adopt a stochastic model of returns and other quantities,
$\hat Z$ could be the conditional
expectation of $Z$, given all data that is available
at the beginning of period $t$, when the asset trades are chosen.

Before proceeding we note that most of the effort in developing a
good trading algorithm goes
into forming the estimates or forecasts, especially of the return $r_t$
\cite{campbell1997econometrics,grinold1999active}.  In this paper, however, we consider
the estimates as given.  Thus we focus on the question, given a set
of estimates, what is a good way to trade based on them?
Even though we do not focus on how the estimates should be constructed,
the ideas in this paper are useful in the development of estimates, since
the value of a set of estimates can depend considerably on how they
are exploited, \ie, how the estimates are turned into trades.
To properly assess the value of a proposed set of estimates or forecasts,
we must evaluate them using a realistic simulation with a good
trading algorithm.

We write our \emph{estimated} portfolio return as
\[
\Rphat _t 
= \hat r_t^T w_t + \hat r_t^Tz_t - \trcosthat_t(z_t) - \hldcosthat_t(w_t+z_t),
\]
which is \eqref{e-return}, with the unknown return $r_t$
replaced with the estimate $\hat r_t$.
The estimated active return is
\[
\Rahat _t =
\hat r_t^T (w_t-\wb_t) + \hat r_t^Tz_t - \trcosthat_t(z_t) - \hldcosthat_t(w_t+z_t).
\]
Each of these consists of a term that does not depend on the trades, plus
\BEQ\label{e-comm-obj}
\hat r_t^Tz_t - \trcosthat_t(z_t) - \hldcosthat_t(w_t+z_t),
\EEQ
the return on the trades minus the transaction and holding costs.

\section{Risk-return optimization}
In a basic optimization-based trading strategy,
we determine the normalized asset trades $z_t$ by solving the
optimization problem
\begin{equation}\label{e-spo1}
\begin{array}{ll}
\mbox{maximize} &
\Rphat_t - \gamma_t \psi_t(w_t+z_t) \\
\mbox{subject to} &
z_t \in \mathcal Z_t, \quad
w_t + z_t \in \mathcal W_t \\
& \ones^Tz_t + \trcosthat_t (z_t) + \hldcosthat_t (w_t + z_t) = 0,
\end{array}
\end{equation}
with variable $z_t$.
Here $\psi_t: \reals^{n+1} \to \reals$ is a risk function,
described below, and $\gamma_t >0$ is the \emph{risk aversion parameter}.
The objective in~\eqref{e-spo1} is called the \emph{risk-adjusted estimated return}.
The sets $\mathcal Z_t$ and $\mathcal W_t$
are the trading and holdings constraint sets, respectively,
also described in more detail below.
The current portfolio weight $w_t$ is known, \ie, a parameter, in the problem
\eqref{e-spo1}.
The risk function, constraint sets, and estimated transaction and holding costs
can all depend on the portfolio value $v_t$, but we suppress this dependence
to keep the notation light.

To optimize performance against the risk-free interest rate
or a benchmark portfolio,
we replace $\Rphat_t$ in \eqref{e-spo1} with $\Rehat_t$ or $\Rahat_t$.
By \eqref{e-comm-obj}, these all have the form of a
constant that does not depend on $z_t$, plus
\[
\hat r_t ^T z_t - \trcosthat_t (z_t) - \hldcosthat_t (w_t + z_t).
\]
So in all three cases we get the same trades by solving the problem
\begin{equation}\label{e-spo}
\begin{array}{ll}
\mbox{maximize} &
\hat r_t^T z_t
- \trcosthat_t (z_t) - \hldcosthat_t (w_t + z_t)
-\gamma_t \psi_t(w_t+z_t)\\
\mbox{subject to} &
z_t \in \mathcal Z_t, \quad
w_t + z_t \in \mathcal W_t \\
& \ones^Tz_t + \trcosthat_t (z_t) + \hldcosthat_t (w_t + z_t) = 0,
\end{array}
\end{equation}
with variable $z_t$.
(We will see later that the risk functions are not the same for absolute, excess, and
active return.)
The objective has four terms: The first is the estimated return
for the trades, the second is the estimated transaction cost,
the third term is the holding cost of the post-trade portfolio,
and the last is the risk of the post-trade portfolio.
Note that the first two depend on the trades $z_t$ and the last two
depend on the post-trade portfolio $w_t+z_t$.
(Similarly, the first constraint depends on the trades, and the second on
the post-trade portfolio.)

\paragraph{Estimated versus realized transaction and holding costs.}
The asset trades we choose are given by $(z_t)_{1:n}=(z_t^\star)_{1:n}$,
where $z_t^\star$ is optimal for \eqref{e-spo}.
In dollar terms, the asset trades are $(u_t)_{1:n}=v_t(z_t^\star)_{1:n}$.

The true normalized cash trade value $(z_t)_{n+1}$ is
found by the self-financing
condition \eqref{e-cash-trade-weights} from the non-cash asset trades
$(z^\star_t)_{1:n}$ and the realized costs.
This is not (in general) the same as $(z_t^\star)_{n+1}$,
the normalized cash trade value found by solving the
optimization problem \eqref{e-spo}.
The quantity $(z_t)_{n+1}$ is the normalized cash trade value with the
\emph{realized} costs, while $(z_t^\star)_{n+1}$ is the normalized cash trade value
with the \emph{estimated} costs.

The (small) discrepancy between the realized cash trade value $(z_t)_{n+1}$ and
the planned or estimated cash trade value $(z_t^\star)_{n+1}$
has an implication for the post-trade holding
constraint $w_t+z_t^\star \in \mathcal W_t$.  When we solve \eqref{e-spo}
we require that the post-trade portfolio with the \emph{estimated}
cash balance satisfies
the constraints, which is not quite the same as requiring that
the post-trade portfolio with the \emph{realized} cash balance satisfies
the constraints.  The discrepancy is typically very small, since our estimation
errors for the transaction cost are typically small compared to the
true transactions costs, which in turn are small compared to
the total portfolio value.
But it should be remembered that the realized post-trade portfolio
$w_t + z_t$ can (slightly) violate the constraints since we only constrain
the estimated post-trade portfolio $w_t + z_t^\star$ to satisfy the constraints.
(Assuming perfect trade execution, constraints relating to the
asset portion of the post-trade portfolio
$(w_t+z_t^\star)_{1:n}$ will hold exactly.)

\paragraph{Simplifying the self-financing constraint.}
We can simplify problem \eqref{e-spo} by replacing the self-financing constraint
\[
\ones^Tz_t + \trcosthat_t (z_t) + \hldcosthat_t (w_t + z_t) = 0
\]
with the constraint $\ones^Tz_t = 0$.
In all practical cases, the cost terms are small compared to the total portfolio
value, so the approximation is good.
At first glance it appears that by using the simplified constraint $\ones^Tz=0$
in the optimization problem, we are essentially ignoring the transaction
and holding costs, which
would not produce good results.
But we still take the transaction and holding costs into account in the objective.

With this approximation we obtain the simplified problem
\begin{equation}\label{e-spo-nsf}
\begin{array}{ll}
\mbox{maximize} &
\hat r_t^T z_t
- \trcosthat_t (z_t) - \hldcosthat_t (w_t + z_t)
-\gamma_t \psi_t(w_t+z_t)\\
\mbox{subject to} &
\ones^Tz_t = 0, \quad z_t \in \mathcal Z_t, \quad
w_t + z_t \in \mathcal W_t.
\end{array}
\end{equation}
The solution $z_t^\star$ to the simplified problem
slightly over-estimates the realized cash trade $(z_t)_{n+1}$,
and therefore the post-trade cash balance $(w_t+z_t)_{n+1}$.
The cost functions used
in optimization are only estimates of what the realized values will be;
in most practical cases this estimation error is much larger than
the approximation introduced with the simplification $\ones^Tz_t=0$.
One small advantage (that will be useful in the multi-period trading case)
is that in the optimization problem \eqref{e-spo-nsf}, $w_t+z_t$ is a bona fide
set of weights, \ie, $\ones^T(w_t+z_t)=1$;
whereas in \eqref{e-spo},
$\ones^T(w_t+z_t)$ is (typically) slightly less than one.

We can re-write the problem \eqref{e-spo-nsf} in terms of
the variable $w_{t+1} = w_t +z_t$, which we interpret as the post-trade
portfolio weights:
\begin{equation}\label{e-spo-wt}
\begin{array}{ll}
\mbox{maximize} &
\hat r_t^T w_{t+1}
- \trcosthat_t (w_{t+1}-w_t) - \hldcosthat_t (w_{t+1})
-\gamma_t \psi_t(w_{t+1})\\
\mbox{subject to} &
\ones^Tw_{t+1} = 1, \quad w_{t+1}-w_t \in \mathcal Z_t, \quad
w_{t+1} \in \mathcal W_t,
\end{array}
\end{equation}
with variable $w_{t+1}$.

\section{Risk measures}
\label{s-risk-measures}
The risk measure $\psi_t$ in \eqref{e-spo} or \eqref{e-spo-nsf}
is traditionally an estimate of the variance of the return,
using a stochastic model of the returns \cite{markowitz1952portfolio, kolm2014years}.
But it can be any function that measures our perceived
risk of holding a portfolio.
We first describe the traditional risk measures.

\paragraph{Absolute risk.}
Under the assumption that the returns $r_t$ are stochastic, with
covariance matrix $\Sigma_t \in \reals^{(n+1)\times (n+1)}$,
the variance of $\Rp_t$ is given by
\[
\var(\Rp_t) = (w_t + z_t)^T \Sigma_t (w_t +z_t ).
\]
This gives the traditional quadratic risk measure for period $t$,
\[
\psi_t(x) = x^T \Sigma_t x.
\]
It must be emphasized that $\Sigma_t$ is an \emph{estimate} of the
return covariance under the assumption that the returns are stochastic.
It is usually assumed that the cash return (risk-free interest rate)
$(r_t)_{n+1}$ is known, in which case the last row and column of $\Sigma_t$ are zero.

\paragraph{Active risk.}
With the assumption that $r_t$ is stochastic with covariance $\Sigma_t$,
the variance of the active return $\Rap_t$ is
\[
\var(\Rap_t) = (w_t + z_t - \wb_t)^T\Sigma_t(w_t+z_t-\wb_t).
\]
This gives the traditional quadratic active risk measure
\[
\psi_t(x) = (x-\wb_t)^T\Sigma_t(x-\wb_t).
\]
When the benchmark is cash, this reduces to $x^T\Sigma_t x$, the absolute risk,
since the last row and column of $\Sigma_t$ are zero.
In the sequel we will work with the active risk,
which reduces to the absolute or excess risk when the benchmark is cash.

\paragraph{Risk aversion parameter.}
The risk aversion parameter $\gamma_t$ in \eqref{e-spo} or \eqref{e-spo-nsf}
is used to scale the relative importance of the estimated return and the estimated
risk.  Here we describe how the particular value $\gamma_t=1/2$
arises in an approximation of maximizing expected growth rate, neglecting
costs.
Assuming that the returns $r_t$ are independent samples from a distribution,
and $w$ is fixed, the portfolio return $\Rp_t =w^Tr_t$ is a (scalar) random variable.
The weight vector that maximizes the expected portfolio growth rate
$\Expect \log (1+ \Rp_t)$ (subject to $\ones^Tw=1$, $w \geq 0$) is called
the \emph{Kelly optimal portfolio} or \emph{log-optimal portfolio} \cite{kelly1956new,busseti2016risk}.
Using the quadratic approximation of the logarithm
$\log(1+a) \approx a- (1/2)a^2$ we obtain
\BEAS
\Expect \log (1+\Rp_t) &\approx& \Expect \left( \Rp_t - (1/2) (\Rp_t)^2 \right)\\
&=& \mu^T w - (1/2) w^T(\Sigma+\mu\mu^T) w,
\EEAS
where $\mu=\Expect r_t$ and $\Sigma=\Expect(r_t-\mu)(r_t-\mu)^T$
are the mean and covariance of the return $r_t$.
Assuming that the term $\mu \mu^T$ is small compared to $\Sigma$
(which is the case for realistic daily returns and covariance),
the expected growth rate can be well approximated as
$\mu^T w -(1/2) w^T \Sigma w$.
So the choice of risk aversion parameter $\gamma_t = 1/2$
in the single-period optimization problems
\eqref{e-spo} or \eqref{e-spo-nsf}
corresponds to approximately maximizing growth rate, \ie,
Kelly optimal trading.   In practice it is found that
Kelly optimal portfolios tend to have too much risk \cite{busseti2016risk},
so we expect that useful values of
the risk aversion parameter $\gamma_t$ are bigger than $1/2$.

\paragraph{Factor model.}
When the number of assets $n$ is large,
the covariance estimate $\Sigma_t$ is typically specified
as a low rank (`factor') component, plus a diagonal matrix,
\[
\Sigma_t = F_t \Sigma_t^\mathrm{f} F_t^T + D_t,
\]
which is called a \emph{factor model} (for quadratic risk).
Here $F_t \in \reals^{(n+1)\times k}$ is the \emph{factor loading matrix},
$\Sigma_t^\mathrm{f} \in \reals^{k \times k}$ is an estimate of the covariance of $F^Tr_t$
(the vector of \emph{factor returns}),
and $D_t \in \reals^{(n+1)\times (n+1)}$ is a nonnegative diagonal matrix.

The number of factors $k$ is much less than $n$ (typically, tens versus thousands).
Each entry $(F_t)_{ij}$ is the loading (or \emph{exposure}) of asset $i$ to factor $j$.
Factors can represent economic concepts such as industrial sectors,
exposure to specific countries, accounting measures, and so on.
For example, a technology factor would have loadings of 1 for technology assets
and 0 for assets in other industries.
But the factor loading matrices can be found using many other methods,
for example by a purely data-driven analysis.
The matrix $D_t$ accounts for the additional variance in individual asset returns
beyond that predicted by the factor model, known as the
\emph{idiosyncratic risk}.

When a factor model is used in the problems \eqref{e-spo} or \eqref{e-spo-nsf},
it can offer a very substantial increase in the speed of solution
\cite{perold1984large,boyd2004convex}.
Provided the problem is formulated in such a way that the solver
can exploit the factor model, the computational complexity drops from
$O(n^3)$ to $O(nk^2)$ flops, for a savings of $O((n/k)^2)$.
The speedup can be substantial when
(as is typical) $n$ is on the order of thousands
and $k$ on the order of tens.
(Computational issues are discussed in more detail in \S\ref{s-convexity}.)

We now mention some less traditional risk functions that can be very useful
in practice.

\paragraph{Transformed risk.}
We can apply a nonlinear transformation to the usual quadratic risk,
\[
\psi_t(x) = \varphi((x-\wb_t)^T \Sigma_t (x-\wb_t)),
\]
where $\varphi:\reals \to \reals$ is a nondecreasing function.
(It should also be convex, to keep the optimization problem
tractable, as we will discuss below.)
This allows us to shape our aversion to different levels of
quadratic risk. For example, we can take $\varphi(x) = (x-a)_+$.  In this case
the transformed risk assesses no cost for quadratic risk levels up to $a$.
This can be useful to hit a target risk level, or to be maximally aggressive
in seeking returns, up to the risk threshold $a$.
Another option is $\varphi(x) = \exp (x/\eta)$, where $\eta>0$ is a parameter.
This assesses a strong cost to risks substantially larger than $\eta$, and is
closely related to risk aversion used in stochastic optimization.

The solution of the optimization problem \eqref{e-spo} with transformed
risk is the same as the solution with the traditional risk function, but with a
different value of the risk aversion parameter.  So we can think of transformed
risk aversion as a method to automatically tune the risk aversion parameter,
increasing it as the risk increases.

\paragraph{Worst-case quadratic risk.}
We now move beyond the traditional quadratic risk to create a
risk function that is more robust to unpredicted
changes in market conditions. We define the
\emph{worst-case} risk for portfolio $x$ as
\[
\psi_t(x) = \max_{i = 1, \ldots, M} (x-\wb_t)^T\Sigma^{(i)}_t (x-\wb_t).
\]
Here $\Sigma^{(i)}$, $i=1,\ldots, M$,
are $M$ given covariance matrices;
we refer to $i$ as the \emph{scenario}.
We can motivate the worst-case risk
by imagining that the returns are generated from one of $M$ distributions,
with covariances $\Sigma^{(i)}$ depending on which scenario occurs.
In each period, we do not know, and do not attempt to predict, which scenario
will occur.
The worst-case risk is the largest risk under the $M$ scenarios.

If we estimate the probabilities of occurrence of
the scenarios, and weight the scenario covariance matrices by these probabilities,
we end up back with a single quadratic risk measure,
the weighted sum of the scenario covariances.
It is critical that we combine them using the maximum, and not a
weighted sum. (Although other nonlinear combining functions would also work.)
We should think of the scenarios as describing situations that could arise,
but that we cannot or do not attempt to predict.

The scenario covariances $\Sigma^{(i)}$ can be found by many reasonable methods.
They can be empirical covariances estimated from realized (past) returns conditioned on the
scenario, for example, high or low market volatility, high or low interest rates,
high or low oil prices, and so on \cite{meucci2010historical}.
They could be an analyst's best guess for what the asset covariance would be
in a situation that could occur.


\section{Forecast error risk}
The risk measures considered above attempt to model the period to period
variation in asset returns, and the associated period to period variation in the
portfolio return they induce.
In this section we consider terms that take into account errors in
our prediction of return and covariance.
(The same ideas can be applied to other parameters that we estimate, like
volume.)
Estimation errors can significantly impact the resulting portfolio
weights, resulting in poor out-of-sample performance
\cite{jorion1985international,michaud1989markowitz,chopra1993effect,
kan2007optimal,demiguel2009optimal,fabozzi2010robust,kolm2014years}.

\paragraph{Return forecast error risk.}
We assume our forecasts of the return vector $\hat r$
are uncertain: Any forecast $\hat r + \delta$ with $|\delta| \leq \rho$
and $\rho \in \reals^n$ is possible and consistent with what we know.
In other words,
$\rho$ is a vector of uncertainties on our return prediction $\hat r$.
If we are confident in our (nominal) forecast of the return of asset $i$, we take
$\rho_i$ small; conversely large $\rho_i$ means that we are not
very confident in our forecast.
The uncertainty in return forecast is readily interpreted when
annualized; for example, our uncertain return forecast for an asset might be
described as $6\% \pm 2\%$, meaning any forecast return between
$4\%$ and $8\%$ is possible.

The post-trade estimated return is then
$(\hat r_t + \delta_t)^T (w_t+z_t)$;
we define the minimum of this over $|\delta| \leq \rho$ as the
\emph{worst-case return forecast}.  It is easy to see
what the worst-case value of $\delta$ is: If we hold a long position,
the return (for that asset) should take its minimum value $\hat r_i + \rho_i$;
if we hold a short position, it should take its
maximum allowed value $\hat r_i - \rho_i$.
The worst-case return forecast has the value
\[
\hat R^\mathrm{wc}_t = \hat r_t^T(w_t + z_t-\wb_t) - \rho^T |w_t + z_t-\wb_t|.
\]
The first term here is our original estimate
(including the constant terms we neglect in \eqref{e-spo} and \eqref{e-spo-nsf});
the second term (which is always
nonpositive) is the worst possible value of our estimated active return over
the allowed values of $\delta$.  It is a risk associated with forecast
uncertainty.
This gives
\BEQ\label{e-return-forecast-risk}
\psi_t (x) = \rho^T |x-\wb_t|.
\EEQ
(This would typically be added to a traditional quadratic risk measure.)
This term is a weighted $\ell_1$-norm of the deviation from the weights,
and encourages weights that deviate sparsely from the benchmark,
\ie, weights with some or many entries equal to those of the
benchmark \cite{fastrich2015constructing,ho2015weighted,li2015sparse}.

\paragraph{Covariance forecast error risk.}
In a similar way we can add a term that corresponds to risk of errors in forecasting
the covariance matrix in a traditional quadratic risk model.
As an example, suppose that we are given a nominal covariance matrix $\Sigma$,
and consider the \emph{perturbed covariance matrix}
\[
\Sigma^\mathrm{pert} = \Sigma + \Delta,
\]
where $\Delta$ is a symmetric perturbation matrix with
\BEQ\label{e-pert-limit}
|\Delta_{ij} | \leq \kappa \left(\Sigma_{ii}\Sigma_{jj}\right)^{1/2},
\EEQ
where $\kappa \in [0,1)$ is a parameter.
This perturbation model means that the diagonal entries of covariance can
change by the fraction $\kappa$; ignoring the change in the diagonal entries,
the asset correlations can change by up to (roughly) $\kappa$.  The value of
$\kappa$ depends on our confidence in the covariance matrix; reasonable values
are $\kappa = 0.02$, $0.05$, or more.

With $v=x-\wb_t$, the maximum (worst-case) value of the quadratic risk over
this set of perturbations is given by
\BEAS
\max_{|\Delta_{ij}| \leq \kappa (\Sigma_{ii}\Sigma_{jj})^{1/2} }
v^T (\Sigma^\mathrm{pert})v &=&
\max_{|\Delta_{ij}| \leq \kappa (\Sigma_{ii}\Sigma_{jj})^{1/2} }
v^T (\Sigma + \Delta) v\\ &=&
v^T \Sigma v + \max_{|\Delta_{ij}| \leq \kappa (\Sigma_{ii}\Sigma_{jj})^{1/2} }
\sum_{ij} v_i v_j \Delta_{ij}\\
&=&
v^T \Sigma v + \kappa \sum_{ij} |v_i v_j| (\Sigma_{ii}\Sigma_{jj})^{1/2} \\
&=&
v^T \Sigma v + \kappa \left( \sum_i \Sigma_{ii}^{1/2} |v_i|\right)^2.
\EEAS
This shows that the worst-case covariance,
over all perturbed covariance matrices consistent with our
risk forecast error assumption \eqref{e-pert-limit}, is given by
\BEQ\label{e-risk-forecast-risk}
\psi_t (x)= (x-\wb_t)^T \Sigma (x-\wb_t) +
\kappa \left(\sigma^T |x-\wb_t|\right)^2,
\EEQ
where $\sigma = (\Sigma_{11}^{1/2}, \ldots, \Sigma_{nn}^{1/2})$ is the vector of
asset volatilities.
The first term is the usual quadratic risk with the nominal
covariance matrix; the second term can be interpreted as risk associated
with covariance forecasting error \cite{ho2015weighted,li2015sparse}.
It is the square of a weighted $\ell_1$-norm of the deviation
of the weights from the benchmark.
(With cash benchmark, this directly penalizes large leverage.)


\section{Holding constraints}
\label{s-holding-constr}
Holding constraints restrict our choice
of normalized post-trade portfolio $w_t + z_t$.
Holding constraints may be surrogates for constraints on $w_{t+1}$,
which we cannot constrain directly since it depends on the unknown returns.
Usually returns are small and $w_{t+1}$ is close to $w_t + z_t$,
so constraints on $w_t + z_t$ are good approximations for constraints on $w_{t+1}$.
Some types of constraints always hold exactly for $w_{t+1}$ when they hold for
$w_t + z_t$.

Holding constraints may be mandatory, imposed by law or the investor,
or discretionary, included to avoid certain undesirable portfolios.
We discuss common holding constraints below.
Depending on the specific situation, each of these constraints could be imposed
on the \emph{active} holdings $w_t + z_t - \wb_t$ instead of the absolute
holdings $w_t + z_t$, which we use here for notational simplicity.

\paragraph{Long only.}
This constraint requires that only long asset positions are held,
\[
w_t + z_t \geq 0.
\]
If only the assets must be long, this becomes $(w_t + z_t)_{1:n} \geq 0$.
When a long only constraint is imposed on the post-trade weight $w_t+z_t$,
it automatically holds on the next period value $(\ones+r_t)\circ(h_t+z_t)$,
since $\ones + r_t \geq 0$.

\paragraph{Leverage constraint.}
The leverage can be limited with the constraint
\[
\|(w_t + z_t)_{1:n} \|_1 \leq L^\mathrm{max},
\]
which requires the post-trade portfolio leverage to not exceed $L^\mathrm{max}$.
(Note that the leverage of the next period portfolio can be slightly
larger than $L^\mathrm{max}$, due to the returns over the period.)

\paragraph{Limits relative to asset capitalization.}
Holdings are commonly limited so that the investor does not own
too large a portion of the company total value.
Let $C_t$ denote the vector of asset capitalization, in dollars.
The constraint
\[
(w_t + z_t)_i \leq \delta \circ C_t/v_t,
\]
where $\delta \geq 0$ is a vector of fraction limits, and $/$ is
interpreted elementwise,
limits the long post-trade position
in asset $i$ to be no more than the fraction $\delta_i$ of the capitalization.
We can impose a similar limit on short positions, relative to
asset capitalization, total outstanding short value, or some combination.

\paragraph{Limits relative to portfolio.}
We can limit our holdings in each asset to lie between a minimum and
a maximum fraction of the portfolio value,
\[
-w^\mathrm{min} \leq w_t+z_t \leq w^\mathrm{max},
\]
where $w^\mathrm{min}$ and $w^\mathrm{max}$ are nonnegative
vectors of the maximum short and long allowed fractions, respectively.
For example with $w^\mathrm{max}=w^\mathrm{min} = (0.05) \ones$,
we are not allowed to hold more than 5\% of the portfolio value in any
one asset, long or short.

\paragraph{Minimum cash balance.}
Often the cash balance must stay above a minimum dollar threshold $c_\mathrm{min}$
(which can be negative).
We express a minimum cash balance as the constraint
\[
(w_t + z_t)_{n+1} \geq c_\mathrm{min}/v_t.
\]
This constraint can be slightly violated by the realized values, due to
our error in estimation of the costs.

\paragraph{No-hold constraints.}
A no-hold constraint on asset $i$ forbids holding a position in asset $i$,
\ie,
\[
(w_t + z_t)_i = 0.
\]

\paragraph{$\beta$-neutrality.}
A $\beta$-neutral portfolio is one whose return $\Rp$
is uncorrelated with the benchmark return $\Rb$, according to our
estimate $\Sigma_t$ of $\cov(r_t)$.
The constraint that $w_t + z_t$ be $\beta$ neutral takes the form
\[
(w^\mathrm{b}_t)^T \Sigma_t (w_t + z_t) = 0.
\]

\paragraph{Factor neutrality.}
In the factor covariance model, the estimated
portfolio risk $\sigma^\mathrm{F}_i$ due to factor $i$ is given by
\[
\left( \sigma^\mathrm{F}_i\right)^2
= (w_t + z_t)^T (F_t)_i (\Sigma^\text{f}_t)_{ii} (F_t)_i^T (w_t + z_t).
\]
The constraint that the portfolio be neutral to factor $i$ means that
$\sigma^\mathrm{F}_i = 0$,
which occurs when
\[
(F_t)_i^T (w_t + z_t) = 0.
\]

\paragraph{Stress constraints.}

Stress constraints protect the portfolio against unexpected changes in market
conditions.
Consider scenarios $1,\ldots,K$, each representing a market shock event
such as a sudden change in oil prices,
a general reversal in momentum, or a collapse in real estate prices.
Each scenario $i$ has an associated (estimated) return $c_i$.
The $c_i$ could be based on past occurrences of scenario $i$
or predicted by analysts
if scenario $i$ has never occurred before.

Stress constraints take the form
\[
c_i^T(w_t + z_t) \geq R^\mathrm{min},
\]
\ie, the portfolio return in scenario $i$ is above $R^\mathrm{min}$.
(Typically $R^\mathrm{min}$ is negative; here we are limiting the decrease in portfolio
value should scenario $i$ actually occur.)
Stress constraints are related to chance constraints such as value at risk in the sense that
they restrict the probability of large losses due to shocks.



\paragraph{Liquidation loss constraint.}
We can bound the loss of value incurred by liquidating
the portfolio over $T^\mathrm{liq}$ periods.
A constraint on liquidation loss will deter the optimizer from investing in
illiquid assets.
We model liquidation as the transaction cost to trade $h^+$ over
$T^\mathrm{liq}$ periods.
If we use the transaction cost estimate $\hat \phi$ for all periods,
the optimal schedule is to trade $(w_t + z_t)/T^\mathrm{liq}$ each period.
The constraint that the liquidation loss is no more than the fraction $\delta$
of the portfolio value is given by
\[
T^\mathrm{liq} \trcosthat_t((w_t + z_t)/T^\mathrm{liq}) \leq \delta.
\]
(For optimization against a benchmark, we replace this with the
cost to trade the portfolio to the benchmark
over $T^\mathrm{liq}$ periods.)

\paragraph{Concentration limit.}
As an example of a non-traditional constraint, we consider a
\emph{concentration limit}, which requires that no more than a given fraction $\omega$
of the portfolio value can be held in some given fraction
(or just a specific number $K$) of assets.  This can be written as
\[
\sum_{i=1}^K (w_t+z_t)_{[i]} \leq \omega,
\]
where the notation $a_{[i]}$ refers to the $i$th largest
element of the vector $a$.  The left-hand side is the sum of the $K$ largest
post-trade positions.
For example with $K=20$ and $\omega = 0.4$, this constraint prohibits holding
more than $40\%$ of the total value in any $20$ assets.
(It is not well known that this constraint is convex, and indeed,
easily handled; see \cite[\S 3.2.3]{boyd2004convex}.
It is easily extended to the case where $K$ is not an integer.)

\section{Trading constraints}
\label{s-trading-constr}

Trading constraints restrict the choice of normalized trades $z_t$.
Constraints on the non-cash trades $(z_t)_{1:n}$
are exact (since we assume that our trades are executed in full),
while constraints on the cash trade $(z_t)_{n+1}$ are approximate,
due to our estimation of the costs.
As with holding constraints, trading constraints may be mandatory
or discretionary.

\paragraph{Turnover limit.}
The turnover of a portfolio in period $t$ is given by $\|(z_t)_{1:n}\|_1/2$.
It is common to limit the turnover to a fraction $\delta$ (of portfolio value),
\ie,
\[
\|(z_t)_{1:n}\|_1/2 \leq \delta.
\]

\paragraph{Limits relative to trading volume.}
Trades in non-cash assets may be restricted to a certain fraction
$\delta$ of the current period market volume $V_t$ (estimate),
\[
|(z_t)_{1:n}| \leq \delta (V_t/v_t),
\]
where the division on the right-hand side means elementwise.

\paragraph{No-buy, sell, or trade restriction.}
A no-buy restriction on asset $i$ imposes the constraint
\[
(z_t)_i \leq 0,
\]
while a no-sell restriction imposes the constraint
\[
(z_t)_i \geq 0.
\]
A no-trade restriction imposes both a no-buy and no-sell restriction.

\section{Soft constraints}
Any of the constraints on holdings or transactions can be made \emph{soft},
which means that are not strictly enforced.
We explain this in a general setting.
For a vector equality constraint $h(x)=0$ on the variable or expression $x$, we replace
it with a term subtracted the objective of the form
$\gamma \|h(x)\|_1$,
where $\gamma > 0$ is the \emph{priority} of the constraint.
(We can generalize this to $\gamma^T |h(x)|$, with $\gamma$ a vector, to give
different priorities to the different components of $h(x)$.)
In a similar way we can replace an inequality constraint $h(x)\leq 0$ with
a term, subtracted from the objective, of the form $\gamma^T(h(x))_+$,
where $\gamma >0$ is a vector of priorities.
Replacing the hard constraints with these \emph{penalty} terms results in
\emph{soft constraints}.  For large enough values of the priorities, the
constraints hold exactly; for smaller values, the constraints are (roughly speaking)
violated only when they need to be.

As an example, we can convert a set of factor neutrality constraints
$F_t^T (w_t+z_t)=0$ to soft constraints,
by subtracting a term $\gamma \| F_t^T (w_t+z_t) \|_1$ from the objective,
where $\gamma>0$ is the priority.
For larger values of $\gamma$ factor neutrality $ F_t^T (w_t+z_t)=0$
will hold (exactly, when possible); for smaller
values some factor exposures can become nonzero, depending on other
objective terms and constraints.

\section{Convexity}\label{s-convexity}
The portfolio optimization problem \eqref{e-spo}
can be solved quickly and reliably
using readily available software so long as the problem is convex.
This requires that the risk and estimated transaction
and holding cost functions are convex, and the trade and holding constraint
sets are convex.
All the functions and constraints discussed above are convex,
except for the self-financing constraint
\[
\ones^Tz_t + \trcosthat_t(z_t) + \hldcosthat_t(w_t + z_t) = 0,
\]
which must be relaxed to the inequality
\[
\ones^Tz_t + \trcosthat_t(z_t) + \hldcosthat_t(w_t + z_t) \leq 0.
\]
The inequality will be tight at the optimum of \eqref{e-spo}.
Alternatively, the self-financing constraint can be replaced with the
simplified version $\ones^Tz_t = 0$ as in problem \eqref{e-spo-nsf}.

\paragraph{Solution times.}
The SPO problems described above, except for the multi-covariance risk model,
can be solved using standard interior-point methods with a complexity $O(nk^2)$
flops, where $n$ is the number of assets and $k$ is the number of factors.
(Without the factor model, we replace $k$ with $n$.)
The coefficient in front is on the order of $100$, which includes the
interior-point iteration count and other computation.
This should be the case even for complex leverage constraints,
the $3/2$ power transaction costs, limits on trading and holding, and so on.

This means that a typical current single core (or thread) of a processor can solve an
SPO problem with 1500 assets and 50 factors in under one half second
(based conservatively on a computation speed of 1G flop/sec).
This is more than fast enough to use the methods to carry out trading with
periods on the order of a second.  But the speed is still very
important even when the trading is daily, in order to carry out back-testing.
For daily trading, one year of back-testing, around 250 trading days, can be
carried out in a few minutes or less.  A generic 32 core computer, running 64 threads,
can carry out a back-test on five years of data, with 64 different choices of
parameters (see below), in under 10 minutes.  This involves solving 80000
convex optimization problems.
All of these times scale linearly with the number of assets, and quadratically with
the number of factors.  For a problem with, say, 4500 assets and 100 factors,
the computation times would be around $12 \times$ longer.
Our estimates are conservatively based on a computation speed of 1G flop/sec;
for these or larger problems multi-threaded optimized linear algebra routines can
achieve 100G flop/sec, making the back-testing $100 \times$ faster.

We mention one special case that can be solved much faster.  If the objective
is quadratic, which means that the risk and costs are quadratic functions,
and the only constraints are linear equality constraints (\eg,
factor neutrality), the problem can be solved with the same $O(nk^2)$ complexity,
but the coefficient in front is closer to $2$, around 50 times faster than
using an interior-point method.

Custom solvers, or solvers targeted to specific platforms like GPUs,
can solve SPO problems much faster.
For example the first order operator-splitting method implemented in
POGS \cite{fougner2015pogs} running on a GPU, can solve extremely large
SPO problems.
POGS can solve a problem with 100000 assets and 1000 factors
(which is much larger than any practical problem) in a few seconds or less.
At the other extreme, code generation systems like CVXGEN \cite{mattingley2012cvxgen}
can solve smaller SPO problems with stunning speed; for example a problem
with 30 assets in well under one millisecond.

\paragraph{Problem specification.}
New frameworks for convex optimization such as
CVX \cite{fougner2015pogs}, CVXPY \cite{cvxpy}, and Convex.jl \cite{convexjl},
based on the idea of disciplined convex programming (DCP) \cite{grant2006dcp},
make it very easy to specify and modify the SPO problem in just a handful
of lines of easy to understand code.   These frameworks make it
easy to experiment with non-standard trading and holding constraints,
or risk and cost functions.

\paragraph{Nonconvexity.}
The presence of nonconvex constraints or terms in the optimization problem
greatly complicates its solution, making its solution time much longer,
and sometimes very much longer.  This may not be a problem in the production
trading engine that determines one trade per day, or per hour.
But nonconvexity makes back-testing much slower
at the least, and in many cases simply impractical.
This greatly reduces the effectiveness of the whole optimization-based approach.
For this reason, nonconvex constraints or terms should be strenuously avoided.

Nonconvex constraints generally arise only when someone who does not understand
this adds a reasonable sounding constraint, unaware of the trouble he or she is
making.  As an example, consider imposing a minimum trade condition, which states
that if $(z_t)_i$ is nonzero, it must satisfy $|(z_t)_i| \geq \epsilon$, where
$\epsilon >0$.  This constraint seems reasonable enough,
but makes the problem nonconvex.
If the intention was to achieve sparse trading, or to avoid
many very small trades, this can be accomplished
(in a far better way) using convex constraints or cost terms.

Other examples of nonconvex constraints (that should be avoided) include
limits on the number of assets held,
minimum values of nonzero holdings, or restricting trades to be integer
numbers of share lots, or restricting the total number of assets we can trade.
The requirement that we must trade integer numbers of shares is also
nonconvex, but irrelevant for any practical portfolio.
The error induced by rounding our trade lists (which contain real numbers) to
an integer number of shares is negligible for reasonably sized portfolios.

While nonconvex constraints and objective terms should be avoided, and
are generally not needed,
it is possible to handle many of them using simple powerful heuristics,
such as solving a relaxation, fixing the nonconvex terms, and then solving
the convex problem again \cite{diamondtakapoui2016}.
As a simple example of this approach,
consider the minimum nonzero trade requirement
$|(z_t)_i| \geq \epsilon$ for $(z_t)_i \neq 0$.
We first solve the SPO problem without this constraint, finding a solution
$\tilde z$. We use this tentative trade vector to determine
which entries of $z$ will be zero, negative, or positive (\ie, which assets
we hold, sell, or buy).  We now impose these
sign constraints on the trade vector:
We require $(z_t)_i=0$ if $(\tilde z_t)_i=0$,
$(z_t)_i\geq 0$ if $(\tilde z_t)_i>0$,
and $(z_t)_i\leq 0$ if $(\tilde z_t)_i<0$.
We solve the SPO again, with these sign
constraints, and the minimum trade constraints as well, which are now
linear, and therefore convex.  This simple method will work very well in
practice.

As another example, suppose that we are limited to make at most $K$ nonzero trades
in any given period.
A very simple scheme, based on convex optimization,  will work extremely well.
First we solve the problem ignoring the limit, and possibly with an additional
$\ell_1$ transaction cost added in, to discourage trading.  We take this trade list
and find the $K$ largest trades (buy or sell).  We then add the constraint to
our problem that we will only trade these assets, and we solve the
portfolio optimization problem again, using only these trades.
As in the example described above, this approach will yield extremely good, if not
optimal, trades.  This approximation will have no effect on the
real metrics of interest, \ie, the portfolio performance.

There is generally no need to solve the nonconvex problem
globally, since this greatly increases the solve time and delivers no
practical benefit in terms of trading performance.
The best method for handling nonconvex problems in portfolio optimization
is to avoid them.

\section{Using single-period optimization}
\label{subsec:using-spo}

\paragraph{The idea.}
In this section we briefly explain, at a high level, how the SPO trading
algorithm is used in practice.
We do not discuss what is perhaps the most critical part,
the return (and other parameter) estimates and forecasts.
Instead, we assume the forecasts are given, and focus on how to use SPO to
exploit them.

In SPO, the parameters that appear in the
transaction and holding costs can be
inspired or motivated by our estimates of what their true values will be,
but it is better to think of them as `knobs' that we turn to
achieve trading behavior that we like
(see, \eg, \cite[chapter 8]{cornuejols2006optimization},
\cite{jagannathan2003risk,demiguel2009generalized,li2015sparse}),
as verified by back-testing, what-if simulation, and stress-testing.

As a crude but important example, we can scale the entire transaction cost
function $\trcost_t$ by a \emph{trading aversion factor} $\gamma^\mathrm{trade}$.
(The name emphasizes the analogy with the risk aversion parameter, which scales
the risk term in the objective.)
Increasing the trading aversion parameter will deter trading or reduce turnover;
decreasing it will increase trading and turnover.
We can even think of $1/\gamma^\mathrm{trade}$ as the number of periods
over which we will amortize the transaction cost we incur \cite{grinold2006dynamic}.
As a more sophisticated example, the transaction cost parameters $a_t$, meant to model
bid-ask spread, can be scaled up or down. If we increase them,
the trades become more sparse, \ie, there are many periods in
which we do not trade each asset.  If we scale the $3/2$-power
term, we encourage or discourage large trades.  Indeed, we could add a
quadratic transaction term to the SPO problem, not because we think it is a
good model of transaction costs, but to discourage large trades even more
than the $3/2$-power term does.  Any SPO variation, such as scaling certain terms,
or adding new ones, is assessed by back-testing and stress-testing.

The same ideas apply to the holding cost.
We can scale the holding cost rates by a positive
\emph{holdings aversion parameter} $\gamma^\mathrm{hold}$
to encourage, or discourage, holding positions that incur
holding costs, such as short positions.
If the holding cost reflects the cost of holding short positions,
the parameter $\gamma^\mathrm{hold}$ scales our aversion to
holding short positions.
We can modify the holding cost by adding a quadratic term of the short
positions $\kappa^T (w_t+z_t)_-^2$,
(with the square interpreted elementwise and $\kappa \geq 0$),
not because our actual borrow cost rates increase with large
short positions, but to send the
message to the SPO algorithm that we wish to avoid holding large short positions.

As another example, we can add a liquidation loss term to the
holding cost, with a scale factor to control its effect.
We add this term not because we intend to liquidate the portfolio, but
to avoid building up large positions in illiquid assets.
By increasing the scale factor for the liquidation loss term, we
discourage the SPO algorithm from holding illiquid positions.
%

\paragraph{Trade, hold, and risk aversion parameters.}
The discussion above suggests that we modify
the objective in~\eqref{e-spo-nsf} with
scaling parameters for transaction and holding
costs, in addition to the traditional risk aversion parameter, which
yields the SPO problem
\BEQ\label{e-spo-hyperparameters}
\begin{array}{ll}
\mbox{maximize} &
\Big(\hat r_t^T z_t
- \gamma_t^\mathrm{trade}\trcosthat_t (z_t) \mbox{} \\
& \qquad \mbox{} - \gamma_t^\mathrm{hold} \hldcosthat_t (w_t + z_t)
 - \gamma_t^\mathrm{risk} \psi_t(w_t+z_t)\Big)\\
\mbox{subject to} &
\ones^Tz_t = 0, \quad z_t \in \mathcal Z_t, \quad
w_t + z_t \in \mathcal W_t.
\end{array}
\EEQ
where $\gamma_t^\mathrm{trade}$,
$\gamma_t^\mathrm{hold}$,
and $\gamma_t^\mathrm{risk}$ are positive parameters used to
scale the respective costs.
These parameters are sometimes called \emph{hyper-parameters},
which emphasizes the analogy to the hyper-parameters used when
fitting statistical models to data.
The hyper-parameters are `knobs' that we `turn' (\ie, choose or change)
to obtain good performance, which we evaluate by back-testing.
We can have even more than three hyper-parameters,
which scale individual terms in the holding and transaction costs.
The choice of hyper-parameters can greatly affect the performance
of the SPO method.  They should be
chosen using back-testing, what-if testing, and stress-testing.


This style for using SPO
is similar to how optimization is used in many other applied areas,
for example control systems or machine learning.   In machine learning,
for example, the goal is to find a model that makes good predictions
on new data.
Most methods for constructing a model use optimization to minimize a
so-called loss function, which penalizes not fitting the observed data,
plus a regularizer, which penalizes model sensitivity or complexity.
Each of these functions is inspired by a (simplistic)
theoretical model of how the data were generated.  But the final choice of
these functions, and the (hyper-parameter)
scale factor between them, is done by out-of-sample
validation or cross validation, \ie, testing the model on data it has not seen.
For general discussion of how convex optimization is used in
this spirit, in applications such as control or estimation,
see \cite{boyd2004convex}.

\paragraph{Judging value of forecasts.}
In this paper we do not consider forecasts, which of course are
critically important in trading.
The most basic test of a new proposed return estimate or forecast is that it
do well predicting returns.  This is typically judged using a simple model
that evaluates Sharpe ratio or information ratio, implicitly ignoring all
portfolio constraints and costs.
If a forecast fails these simple SR or IR tests, it is unlikely to be useful
in a trading algorithm.

But the true value of a proposed estimate or forecast in
the context of multi-period trading can be very different from what is
suggested by the simple SR or IR prediction tests, due to costs,
portfolio constraints, and other issues.
A new proposed forecast should be judged in the context of the portfolio
constraints, other forecasts (say, of volume), transaction costs, holding
costs, trading constraints, and choice of parameters such as risk aversion.
This can be done using simulation,
carrying out back-tests, what-if simulations, and stress-tests,
in each case varying the parameters to achieve the best performance.
The result of this testing is that the forecast might be less valuable
(the usual case) or more valuable (the less usual case) than it appeared
from the simple SR and IR tests.
One consequence of this is that the true value of a forecast can depend
considerably on the type and size of the portfolio being traded; for example, a
forecast could be very valuable for a small long-short portfolio with modest
leverage, and much less valuable for a large long-only portfolio.

\chapter{Multi-Period Optimization}
\label{s-mpo}
\section{Motivation}
In this section we discuss optimization-based strategies that consider
information about multiple periods when choosing trades for the current period.
Before delving into the details, we should consider what we hope to gain
over the single-period approach.
Predicting the returns for the current period is difficult enough.
Why attempt to forecast returns in future periods?

One reason is to better account for transaction costs.
In the absence of transaction cost (and other limitations on trading),
a greedy strategy that only considers one period at a time is optimal,
since performance for the current period does not depend on previous holdings.
However in any realistic model current holdings strongly affect
whether a return prediction can be profitably acted on.
We should therefore consider whether the trades we make in the current period
put us in a good or bad position to trade in future periods.
While this idea can be incorporated into single-period optimization,
it is more naturally handled in multi-period optimization.

For example, suppose our single period optimization-based strategy tells us to
go very long in a rarely traded asset.
We may not want to make the trade because we know that unwinding the position
will incur large transaction costs.
The single-period problem models the cost of moving into the position,
but not the cost of moving out of it.
To model the fact that we will over time revert positions towards the benchmark,
and thus must eventually sell the positions we buy,
we need to model time beyond the current period.
(One standard trick in single-period optimization is to
double the transaction cost, which is then called the
\emph{round-trip cost}.)

Another advantage of multi-period optimization is that it naturally
handles multiple, possibly conflicting return estimates on different
time scales (see, \eg, \cite{garleanu2013dynamic, nystrup2016dynamic}).
As an example, suppose we predict that a return will be positive
over a short period, but over a longer period it will be negative.
The first prediction might be relevant for only a day, while the second
for a month or longer.
In a single-period optimization framework, it is not clear how to account
for the different time scales when blending the return predictions.
Combining the two predictions would likely cancel them, or have us move
according to whichever prediction is larger.
But the resulting behavior could be quite non-optimal.
If the trading cost is high, taking no action is likely
the right choice, since we will have to reverse any trade based on
the fast prediction as we follow the slow prediction in future periods.
If the trading cost is low, however, the right choice is to
follow the fast prediction, since unwinding the position is cheap.
This behavior falls naturally out of a multi-period optimization,
but is difficult to capture in a single-period problem.

There are many other situations where predictions over multiple periods,
as opposed to just the current period, can be taken advantage of in
multi-period optimization.  We describe a few of them here.
\BIT
\item \emph{Signal decay and time-varying returns predictions.}
Generalizing the discussion above on fast versus slow signals,
we may assign an exponential \emph{decay-rate} to every
return prediction signal. (This can be estimated historically,
for example, by fitting an auto-regressive model to the signal values.)
Then it is easy to compute return estimates at any time scale.
The decay in prediction accuracy is also called mean-reversion
or alpha decay (see, \eg, \cite{campbell1997econometrics,grinold2006dynamic,garleanu2013dynamic}).
\item \emph{Known future changes in volatility or risk.}
If we know that a future event will increase the risk, we may want
to exit some of the risky positions in advance.
In MPO, trading
towards a lower risk position starts well before increase in risk,
trading it off with the transaction costs.
In SPO, (larger) trading to a lower risk position occurs only once the
risk has increased, leading to larger transaction costs.
Conversely, known periods of low risk can be exploited as well.
\item  \emph{Changing constraints over multiple periods.}
As an example, assume we want to de-leverage the portfolio over
multiple periods, \ie, reduce the leverage constraint $L^\mathrm{max}$ over some number
of periods to a lower value.
If we use a multi-period optimization framework we will likely
incur lower trading cost than by some ad-hoc approach,
while still exploiting our returns predictions.
\item  \emph{Known future changes in liquidity or volume.}
Future volume or volatility predictions can be exploited for transaction cost optimization,
for example by delaying some trades until they will be cheaper.
Market volumes $V_t$ have much better predictability than market returns.
\item \emph{Setting up, shutting down, or transferring a portfolio.}
These transitions can all be handled naturally by MPO,
with a combination of constraints and objective terms changing over time.
\EIT

\section{Multi-period optimization}
In multi-period optimization, we choose the current trade vector
$z_t$ by solving
an optimization problem over a \emph{planning horizon}
that extends $H$ periods into the future,
\[
t,~t+1, ~ \ldots, ~ t+H-1.
\]
(Single-period optimization corresponds to the case $H=1$.)

Many quantities at times $t,t+1, \ldots, t+H-1$ are unknown at time $t$,
when the optimization problem is solved and the asset trades are chosen,
so as in the single-period case, we will estimate them.
For any quantity or function $Z$, we let $\hat Z_{\tau|t}$ denote our estimate
of $Z_\tau$ given all information available to us at the beginning of period $t$.
(Presumably $\tau \geq t$; otherwise we can take $\hat Z_{\tau|t}=Z_\tau$, the
realized value of $Z$ at time $\tau$.)
For example $\hat r_{t|t}$ is the estimate made at time $t$ of the return at time
$t$ (which we denoted $\hat r_t$ in the section on single-period optimization);
$\hat r_{t+2|t}$ is the estimate made at time $t$ of the return at time $t+2$.

We can develop a multi-period optimization problem
starting from \eqref{e-spo}.
Let
\[
z_t,~ z_{t+1},~ \ldots, ~z_{t+H-1}
\]
denote our sequence of planned trades over the horizon.
A natural objective is the total risk-adjusted return over the
horizon,
\[
\sum_{\tau=t}^{t+H-1} \left( \hat r_{\tau|t}^T (w_\tau + z_\tau)
-\gamma_\tau \psi_\tau(w_\tau+z_\tau)  - \hldcosthat_\tau (w_\tau + z_\tau)
- \trcosthat_\tau (z_\tau)\right).
\]
(This expression drops a constant that does not depend on the trades,
and handles absolute or active return.)
In this expression, $w_t$ is known, but $w_{t+1}, \ldots, w_{t+H}$ are not,
since they depend on the trades $z_t, \ldots, z_{t+H-1}$
(which we will choose) and the unknown returns,
via the dynamics equation \eqref{e-wt1},
\[
w_{t+1} = \frac{1}{1+\Rp_t} (\ones + r_t) \circ (w_t+z_t),
\]
which propagates the current weight vector
to the next one, given the trading and return vectors.
(This true dynamics equation ensures that if $\ones^Tw_t=1$,
we have $\ones^Tw_{t+1}=1$.)

In adding the risk terms $\gamma_\tau \psi_t(w_\tau+z_\tau)$ in this objective,
we are implicitly relying on the idea that the returns are independent
random variables, so the variance of the sum is the sum of the
variances.
We can also interpret $\gamma_\tau \psi_\tau(w_\tau+z_\tau)$ as
cost terms that discourage us from holding certain portfolios.

\paragraph{Simplifying the dynamics.}
We now make a simplifying approximation:  For the purpose of propagating
$w_t$ and $z_t$ to $w_{t+1}$ in our planning exercise,
we will assume $\Rp_t=0$ and $r_t=0$
(\ie, that the one period returns are small compared to one).
This results in the much simpler dynamics equation
$w_{t+1} = w_t + z_t$.
With this approximation, we must add the constraints
$\ones^Tz_t = 0$ to ensure that the weights in our planning
exercise add to one, \ie, $\ones^T w_\tau=1$, $\tau=t+1,\ldots, t+H$.
So we will impose the
constraints
\[
\ones^Tz_\tau = 0, \quad \tau=t+1,\ldots,t+H-1.
\]
The current portfolio weights $w_t$ are given, and satisfy $\ones^T w_t=1$;
we get that $\ones^Tw_\tau = 1$ for $\tau=t+1,\ldots,t+H$ due to the
constraints.
(Implications of the dynamics simplification are discussed below.)

\paragraph{Multi-period optimization problem.}
With the dynamics simplification we arrive at the MPO problem
\begin{equation}\label{e-mpo}
\begin{array}{ll}
\mbox{maximize} &
\sum_{\tau=t}^{t+H-1}
\Big(\hat r_{\tau|t}^T (w_\tau + z_\tau)
-\gamma_\tau \psi_\tau(w_\tau+z_\tau) \mbox{} \\
& \qquad \mbox{} - \hldcosthat_\tau (w_\tau + z_\tau)
- \trcosthat_\tau (z_\tau) \Big) \\
\mbox{subject to} &
\ones^Tz_\tau = 0, \quad
z_\tau \in \mathcal Z_\tau, \quad
w_\tau + z_\tau \in \mathcal W_\tau, \\
& w_{\tau+1} = w_\tau + z_\tau,
\quad \tau = t,\ldots, t+H-1,
\end{array}
\end{equation}
with variables $z_t, z_{t+1}, \ldots, z_{t+H-1}$ and
$w_{t+1}, \ldots, w_{t+H}$.
Note that $w_t$ is not a variable, but the (known) current portfolio weights.
When $H=1$, the multi-period problem reduces to the simplified
single-period problem \eqref{e-spo-nsf}.
(We can ignore the constant $\hat r_{t|t}^T w_t$, which does not
depend on the variables, that appears
in \eqref{e-mpo} but not \eqref{e-spo-nsf}.)

Using $w_{\tau+1}=w_\tau+z_\tau$
we can eliminate the trading variables $z_\tau$ to obtain the equivalent
problem
\begin{equation}\label{e-mpo2}
\begin{array}{ll}
\mbox{maximize} &
\sum_{\tau=t+1}^{t+H} \Big(\hat r_{\tau|t}^T w_\tau
-\gamma_\tau \psi_\tau(w_\tau) \mbox{}\\
& \qquad \mbox{}  - \hldcosthat_\tau (w_\tau)
- \trcosthat_\tau (w_\tau-w_{\tau-1}) \Big) \\
\mbox{subject to} & \ones^Tw_\tau=1, \quad
w_\tau - w_{\tau-1} \in \mathcal Z_\tau,
\quad w_\tau \in \mathcal W_\tau,\\
& \qquad \tau = t+1, \ldots, t+H,
\end{array}
\end{equation}
with variables $w_{t+1}, \ldots, w_{t+H}$, the planned weights
over the next $H$ periods.
This is the multi-period analog of \eqref{e-spo-wt}.

Both MPO formulations \eqref{e-mpo} and \eqref{e-mpo2} are convex optimization
problems, provided the transaction cost, holding cost, risk functions,
and trading and holding constraints are all convex.

\paragraph{Interpretation of MPO.}
The MPO problems \eqref{e-mpo} or \eqref{e-mpo2} can be interpreted as follows.
The variables constitute a \emph{trading plan}, \ie, a set of trades to be executed
over the next $H$ periods.   Solving \eqref{e-mpo} or \eqref{e-mpo2}
is forming a trading plan, based on forecasts of critical quantities
over the planning horizon, and some simplifying assumptions.
We do not intend to execute this sequence of trades,
except for the first one $z_t$.
It is reasonable to ask then why we optimize over the future trades
$z_{t+1}, \ldots, z_{t+H-1}$, since we do not intend to execute them.
The answer is simple:  We optimize over them as part of a \emph{planning
exercise}, just to be sure we don't carry out any trades now (\ie, $z_t$)
that will put us in a bad position in the future.
The idea of carrying out a planning exercise, but only executing the current action,
occurs and is used in many fields, such as automatic control
(where it is called model predictive control, MPC, or receding horizon control)
\cite{kwon2006receding, mattingley2011receding},
supply chain optimization \cite{cho2003supply}, and others.
Applications of MPC in finance
include \cite{herzog2007stochastic, boyd2014performance,
bemporad2014dynamic, nystrup2016dynamic}.

\paragraph{About the dynamics simplification.}
Before proceeding let us discuss the simplification of the dynamics equation,
where we replace the exact weight update
\[
w_{t+1} = \frac{1}{1+\Rp_t} (\ones + r_t) \circ (w_t+z_t)
\]
with the simplified version $w_{t+1} = w_t+z_t$, by assuming that $r_t = 0$.
At first glance it appears to be a gross simplification,
but this assumption is only made for the purpose of propagating the portfolio
forward in our planning process;
we do take the returns into account in the first term of our objective.
We are thus neglecting \emph{second-order} terms, and
we cannot be too far off if the per period returns are small compared to one.

In a similar way, adding the constraints $\ones^Tz_\tau=0$ for
$\tau = t+1, \ldots, t+H-1$  suggests that we are ignoring the
transaction and holding costs,
since if $z_\tau$ were a realized trade we would have
$\ones^Tz_\tau = -\trcost_\tau(z_\tau) - \hldcost_\tau(w_\tau + z_\tau)$.
As above, this assumption is only
made for the purpose of propagating our portfolio forward in
our planning exercise; we do take the costs into account in the objective.

\paragraph{Terminal constraints.}
In MPO, with a reasonably long horizon, we can add a terminal (equality) constraint,
which requires the final planned weight to take some specific value,
$w_{t+H}=w^\mathrm{term}$.
A reasonable choice for the terminal portfolio weight is (our estimate of)
the benchmark weight $\wb$ at period $t+H$.

For optimization of absolute or excess return, the terminal
weight would be cash, \ie, $w^\mathrm{term} = e_{n+1}$.
This means that our planning exercise should finish with the portfolio
all cash.  This does not mean we intend to liquidate the portfolio
in $H$ periods; rather, it means we should carry out our planning
as if this were the case.   This will keep us from making the mistake
of moving into what appears, in terms of our returns predictions, to be
an attractive position that it is, however, expensive to unwind.
For optimization relative to a benchmark, the natural terminal constraint
is to be in the (predicted) benchmark.

Note that adding a terminal constraint reduces the number of variables.
We solve the problem \eqref{e-mpo}, but with $w_{t+H}$ a given constant,
not a variable.
The initial weight $w_t$ is also a given constant;
the intermediate weights $w_{t+1}, \ldots, w_{t+H-1}$ are variables.


\section{Computation}\label{s-mpo-computation}
The MPO problem \eqref{e-mpo2} has $Hn$ variables.  In general the complexity
of a convex optimization increases as the cube of the number of variables,
but in this case the special structure of the problem can be exploited
so that the computational effort grows linearly in $H$, the horizon.
Thus, solving the MPO problem \eqref{e-mpo2} should be a factor $H$ slower than
solving the SPO problem \eqref{e-spo-wt}.
For modest $H$ (say, a few tens), this is not a problem.
But for $H=100$ (say) solving the MPO problem can be very challenging.
Distributed methods based on ADMM \cite{boyd2011distributed,boyd2014performance} can be used
to solve the MPO problem using multiple processors.
In most cases we can solve the MPO problem in production.
The issue is back-testing, since we must solve the problem many times,
and with many variations of the parameters.

\section{How MPO is used}
All of the general ideas about how SPO is used apply to MPO as well;
for example, we consider the parameters in the MPO problem as knobs that we adjust
to achieve good performance under back-test and stress-test.
In MPO, we must provide forecasts of each quantity for each period over the
next $H$ periods.  This can be done using sophisticated forecasts, with possibly
different forecasts for each period, or in a very simple way,
with predictions that are constant.

\section{Multi-scale optimization}\label{s-mso}

MPO trading requires estimates of all relevant quantities,
like returns, transaction costs, and risks, over $H$ trading periods
into the future.
In this section we describe a simplification of MPO that requires
fewer predictions, as well as less computation to carry out the
optimization required in each period.
We still create a plan for trades and weights
over the next $H$ periods, but we assume that
trades take place only a few times over the horizon;
in other time periods the planned portfolio is maintained with no trading.
This preserves the idea that we have recourse; but it greatly
simplifies the problem \eqref{e-mpo}.
We describe the idea for three trades, taken in the short term,
medium term, and long term, and an additional
trade at the end to satisfy a terminal constraint $w_{t+H}=\wb$.

Specifically we add the constraint that in \eqref{e-mpo}, trading
(\ie, $z_\tau \neq 0$) only
occurs at specific periods in the future, for
\[
\tau =t, \quad \tau=t+T^\mathrm{med}, \quad \tau=t+T^\mathrm{long},
\quad \tau= t+ H-1,
\]
where
\[
1< T^\mathrm{med} <T^\mathrm{long} < H-1.
\]
We interpret $z^\mathrm{short}= z_t$ as our short term trade,
$z^\mathrm{med} = z_{t+T^\mathrm{med}}$ as our medium term trade,
and $z^\mathrm{long} = z_{t+T^\mathrm{long}}$ as our long term trade,
in our trading plan.  The final nonzero trade $z_{t+H-1}$ is determined
by the terminal constraint.

For example we might take $T^\mathrm{med}=5$ and
$T^\mathrm{long}=21$, with $H=100$.
If the periods represent days,
we plan to trade now (short term), in a week (medium term)
and in month (longer term); in 99 days, we trade to the benchmark.
The only variables we have
are the short, medium, and long term trades,
and the associated weights, given by
\[
w^\mathrm{short} = w_t + z^\mathrm{short}, \quad
w^\mathrm{med} = w^\mathrm{short} + z^\mathrm{med}, \quad
w^\mathrm{long} = w^\mathrm{med} + z^\mathrm{long}.
\]
To determine the trades to make, we solve \eqref{e-mpo} with
all other $z_\tau$ set to zero, and using the weights given above.
This results in an optimization problem with the same form
as \eqref{e-mpo}, but with only three variables each for trading and
weights, and three terms in the objective, plus an additional
term that represents the transaction cost associated with the final
trade to the benchmark at time $t+H-1$.

\chapter{Implementation}
We have developed an open-source Python package
\verb|CVXPortfolio| \cite{cvxportfolio}
that implements the portfolio simulation and optimization concepts discussed in the paper.
The package relies on Pandas \cite{mckinney2012python} for managing data.
Pandas implements structured data types as in-memory databases
(similar to R dataframes) and provides a rich
API for accessing and manipulating them.
Through Pandas, it is easy to couple our package with database backends.
The package uses the convex optimization modeling framework CVXPY \cite{cvxpy}
to construct and solve portfolio optimization problems.

The package provides an object-oriented framework with classes representing
return, risk measures, transaction costs, holding constraints,
trading constraints, etc.
Single-period and multi-period optimization models are constructed
from instances of these classes.
Each instance generates CVXPY expressions and constraints for any given period $t$,
making it easy to combine the instances into a single convex model.
In chapter \ref{s-examples}
we give some simple numerical examples that use \verb|CVXPortfolio|.

\section{Components}
We briefly review the major classes in the software package.
Implementing additional classes, such as novel policies or risk measures, is straightforward.

\paragraph{Return estimates.} Instances of the \verb|AlphaSource| class generate a return
estimate $\hat{r}_{t}$ for period $t$ using only information available at that period.
The simplest \verb|AlphaSource| instance wraps a Pandas dataframe with return estimates
for each period:
\begin{verbatim}
alpha_source = AlphaSource(return_estimates_dataframe)
\end{verbatim}
Multiple \verb|AlphaSource| instances can be blended into a linear combination.

\paragraph{Risk measure.} Instances of the \verb|RiskMeasure| class
generate a convex cost representing a risk measure at a given period $t$.
For example, the \verb|FullSigma| subclass generates the cost $(w_t +
z_t)^T\Sigma_t(w_t + z_t)$ where $\Sigma_t \in \reals^{(n+1) \times (n+1)}$ is
an explicit matrix, whereas the \verb|FactorModel| subclass generates the
cost with a factor model of $\Sigma_t$.
Any risk measure can be switched to absolute or active risk and weighted with a
risk aversion parameter.
The package provides all the risk measures discussed in \S\ref{s-risk-measures}.

\paragraph{Costs.} Instances of the \verb|TcostModel| and \verb|HcostModel| classes generate
transaction and holding cost estimates, respectively.
The same classes work both for modeling costs in a portfolio optimization
problem and calculating realized costs in a trading simulation.
Cost objects can also be used to express other objective terms like soft constraints.

\paragraph{Constraints.} The package provides classes representing each of the
constraints discussed in \S\ref{s-holding-constr} and \S\ref{s-trading-constr}.
For example, instances of the \verb|LeverageLimit| class generate a leverage limit
constraint that can vary by period.
Constraint objects can be converted into soft constraints, which are cost objects.

\paragraph{Policies.} Instances of the \verb|Policy| class take holdings $w_t$ and value $v_t$ and
output trades $z_t$ using information available in period $t$.
Single-period optimization policies are constructed using the
\verb|SinglePeriodOpt| subclass.
The constructor takes an \verb|AlphaSource|, a list of
costs, including risk models (multiplied by their coefficients), and constraints.
For example, the following code snippet constructs a SPO policy:
\begin{verbatim}
spo_policy = SinglePeriodOpt(alpha_source,
                             [gamma_risk*factor_risk,
                             gamma_trade*tcost_model,
                             gamma_hold*hcost_model],
                             [leverage_limit])
\end{verbatim}
Multi-period optimization policies are constructed similarly.
The package also provides subclasses for simple policies such as periodic re-balancing.

\paragraph{Simulator.} The \verb|MarketSimulator| class is used to run trading
simulations, or back-tests.
Instances are constructed with historical returns and other market data, as well
as transaction and holding cost models.
Given a \verb|MarketSimulator| instance \verb|market_sim|, a back-test is run by calling the
\verb|run_backtest| method with an initial portfolio, policy, and start and end
periods:
\begin{verbatim}
backtest_results = market_sim.run_backtest(init_portfolio,
                                           policy,
                                           start_t, end_t)
\end{verbatim}
Multiple back-tests can be run in parallel with different conditions.
The back-test results include all the metrics discussed in chapter \ref{s-metrics}.

\chapter{Examples}\label{s-examples}
In this section we present
simple numerical examples illustrating the ideas developed
above, all carried out using \verb|CVXPortfolio| and
open-source market data (and some approximations where no open
source data is available).
The code for these is available at
\begin{quote}
\mbox{\url{https://github.com/cvxgrp/cvxportfolio/examples}.}
\end{quote}
Given our approximations, and other short-comings of our simulations
that we will mention below,
the particular numerical results we show should not be taken
too seriously.  But the simulations are good enough for us to illustrate
real phenomena, such as the critical role transaction costs
can play, or how important hyper-parameter search can be.

\section{Data for simulation}
\label{s-example-data}
We work with a period of 5 years, from January 2012 through December 2016,
on the components of the S\&P 500 index
as of December 2016.
We select the ones continuously traded in the period.
(By doing this we introduce survivorship bias,
not a concern since these examples have only a didactic purpose.)
We collect open-source market data from
Quandl \cite{QuandlWIKI}.
The data consists of realized daily market returns $r_t$ (computed using
closing prices) and volumes $V_t$.
We use the federal reserve overnight rate for the cash return.
Following \cite{almgren09}
we approximate the daily volatility with a simple estimator,
$(\sigma_t)_i =
|\log (p_t^\text{open})_i -\log (p_t^\text{close})_i|$,
where $(p_t^\text{open})_i$ and
$(p_t^\text{close})_i$ are the open and close prices
for asset $i$ in period $t$.
We could not find open-source data for the bid-ask spread,
so we used the value $a_t = 0.05\%$ (5 basis points) for all
assets and periods.
As holding costs we use
$s_t = 0.01\%$ (1 basis point) for all assets and periods.
We chose standard values for the other parameters
of the transaction and holding
cost models: $b_t = 1$, $c_t = 0$, $d_t = 0$ for all assets and periods.

\section{Portfolio simulation}
To illustrate back-test portfolio simulation,
we consider a portfolio that is meant to track the \emph{uniform portfolio}
benchmark, which has weight $\wb = (\ones/n, 0)$,
\ie, equal fraction of value in all non-cash assets.
This is not a particularly interesting or good benchmark portfolio;
we use it only as a simple example to illustrate the effects
of transaction costs.
The portfolio starts with $w_1 = w^\mathrm{b}$, and due to asset returns
drifts from this weight vector.  We periodically re-balance, which means
using trade vector $z_t = w^\text{b} - w_t$. For other values of $t$
(\ie, the periods in which we do not re-balance) we have $z_t=0$.

We carry out six back-test simulations for each of two initial portfolio
values, \$100M and \$10B.
The six simulations vary in re-balancing frequency:
Daily, weekly, monthly, quarterly, annually, or never (also called `hold'
or `buy-and-hold').
For each simulation we give the portfolio active return $\bar R^\mathrm{a}$
and active risk $\sigma^\mathrm{a}$
(defined in \S\ref{s-relativemetrics}),
the annualized average transaction cost
$
\frac{250}{T}\sum_{t=1}^{T} \phi_t^\text{trade}(z_t),
$
and the annualized average turnover
$
\frac{250}{T}\sum_{t=1}^{T} \|(z_t)_{1:n}\|_1/2.
$

\begin{figure}
\begin{center}
\includegraphics[width=1.0\textwidth]{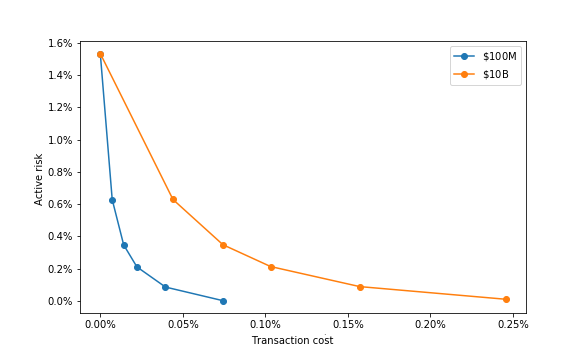}
\end{center}
\caption{
Active risk
versus transaction cost, for the two initial portfolio sizes.
The points on the lines correspond to re-balancing frequencies.
}
\label{fig:basic-frontier}
\end{figure}

Table \ref{tab:basic_example} shows the results.
(The active return is included for completeness,
but has no special meaning; it is the negative of the transaction
cost plus a small random component.)
We observe that transaction cost depends on the total value of
the portfolio, as expected,
and that the choice of re-balancing frequency trades off
transaction cost and active risk.
(The active risk is not exactly zero when
re-balancing daily because of the
variability of the transaction cost, which
is included in the portfolio return.)
Figure \ref{fig:basic-frontier} shows,
separately for the two portfolio sizes,
the active risk versus the transaction cost.

\begin{table}
\begin{center}
\begin{tabular}{ll|llll}
\toprule
Initial    & Rebalancing    & Active  &  Active  &Trans.& Turnover  \\
total val.   & frequency    &  return &  risk&  cost& \\
\midrule
\$100M & Daily &        -0.07\% &       0.00\% &        0.07\% &  220.53\% \\
    & Weekly &        -0.07\% &       0.09\% &        0.04\% &  105.67\% \\
    & Monthly &        -0.12\% &       0.21\% &        0.02\% &   52.71\% \\
    & Quarterly &        -0.11\% &       0.35\% &        0.01\% &   29.98\% \\
    & Annually &        -0.10\% &       0.63\% &        0.01\% &   12.54\% \\
    & Hold &        -0.36\% &       1.53\% &        0.00\% &    0.00\% \\
\$10B & Daily &        -0.25\% &       0.01\% &        0.25\% &  220.53\% \\
    & Weekly &        -0.19\% &       0.09\% &        0.16\% &  105.67\% \\
    & Monthly &        -0.20\% &       0.21\% &        0.10\% &   52.71\% \\
    & Quarterly &        -0.17\% &       0.35\% &        0.07\% &   29.99\% \\
    & Annually &        -0.13\% &       0.63\% &        0.04\% &   12.54\% \\
    & Hold &        -0.36\% &       1.53\% &        0.00\% &    0.00\% \\
\bottomrule
\end{tabular}
\end{center}
\caption{Portfolio simulation results with different initial value
and different re-balancing frequencies. All values are annualized.}
\label{tab:basic_example}
\end{table}

\section{Single-period optimization}
\label{s-example-spo}
In this section we show a simple example of the
single-period optimization model developed in chapter \ref{s-spo}.
The portfolio starts with total value $v_1 =$ \$100M and allocation
equal to the uniform portfolio $w_1 =(\ones/n,0)$.
We impose a leverage constraint of $L^\mathrm{max}=3$.
This simulation uses the market data defined in \S\ref{s-example-data}.
The forecasts and risk model used in the SPO are described below.

\paragraph{Risk model.}
Proprietary risk models, \eg,
from MSCI (formerly Barra), are widely used.
Here we use a simple factor risk model
estimated from past realized returns, using a similar
procedure to \cite{almgren09}.
We estimate it on the first day of each month, and use it for the rest of the month.
Let $t$ be an estimation time period,
and $t-M^\text{risk}$ the time period two years before.
Consider the second moment of the window of realized returns
$\Sigma^\text{exp} = \frac{1}{M^\text{risk}} \sum_{\tau = t-M^\text{risk}}^{t-1}r_{\tau}r_\tau^T$,
and its eigenvalue decomposition $\Sigma^\text{exp} = \sum_{i=1}^n \lambda_i q_i q_i^T$,
where the eigenvalues $\lambda_i$ are in descending order. Our factor risk model is
\[
F=[q_1 \cdots q_k], \quad
\Sigma^\mathrm{f} = \diag(\lambda_1, \ldots, \lambda_k), \quad
D = \sum_{i=k+1}^n \lambda_i \diag( q_i )\diag( q_i ),
\]
with $k=15$.
(The diagonal matrix $D$ is chosen so the factor model $F \Sigma^\mathrm{f} F^T + D$ and
the empirical second moment $\Sigma^\text{exp}$
have the same diagonal elements.)

\paragraph{Return forecasts.}
The risk-free interest rates are known exactly,
$(\hat r_t)_{n+1} = (r_t)_{n+1}$ for all $t$.
Return forecasts for the non-cash assets are always proprietary.
They are generated using many methods,
ranging from analyst predictions to sophisticated machine learning
techniques, based on a variety of data feeds and sources.
For these examples we generate simulated return forecasts
by  adding zero-mean noise to the realized returns
and then rescaling, to obtain return estimates that would (approximately) minimize
mean squared error.
Of course this is not a real return forecast, since it uses the actual realized return;
but our purpose here is only to illustrate the ideas and methods.

For all $t$ the return estimates for non-cash assets are
\BEQ\label{e-SPO-return-forecast}
(\hat r_t)_{1:n} =   \alpha  \left((r_t)_{1:n} + \epsilon_t\right),
\EEQ
where $\epsilon_t \sim \mathcal N(0,\sigma_\epsilon^2 I)$ are independent.
We use noise variance $\sigma^2_\epsilon = 0.02$, so the noise components have
standard deviation around 14\%, around a factor of 10 larger than
the standard deviaton of the realized returns.
The scale factor
$\alpha$ is chosen to minimize the mean squared error
$\Expect ((\hat r_t)_{1:n} - (r_t)_{1:n})^2$,
if we think of $r_t$ as a random variable with variance $\sigma_r$, \ie,
$\alpha = {\sigma^2_r}/{(\sigma^2_r + \sigma^2_\epsilon)}$.
We use the typical value $\sigma^2_r = 0.0005$, \ie, a
realized return standard deviation of around 2\%,
so $\alpha = 0.024$.  Our typical return forecast is on the order of
$\pm 0.3\%$.
This corresponds to an information ratio $\sqrt \alpha \approx 0.15$,
which is on the high end of what might be
expected in practice \cite{grinold1999active}.

With this level of noise and scaling, our return forecasts have an accuracy
on the order of what we might expect from a proprietary forecast.
For example, across all the assets and all days,
the sign of predicted return
agrees with the sign of the real return around 54\% of the times.

\paragraph{Volume and volatility forecasts.}
We use simple estimates of total market volumes and daily volatilities
(used in the transaction cost model),
as moving averages of the realized values with a window of length 10.
For example, the volume forecast at time period $t$ and asset $i$ is
$
(\hat V_t)_i = \frac{1}{10} \sum_{\tau=1}^{10} (V_{t-\tau})_i
$.

\paragraph{SPO back-tests.}
We carry out multiple back-test simulations over the whole period,
varying the risk aversion parameter
$\gamma^\text{risk}$, the trading
aversion parameter $\gamma^\text{trade}$, and the holding cost
multiplier $\gamma^\text{hold}$
(all defined and discussed in \S\ref{subsec:using-spo}).
We first perform a coarse grid search in the hyper-parameter space,
testing all combinations of
\BEAS
\gamma^\text{risk} &=& 0.1,~0.3,~1,~3,~10,~30,~100,~300,~1000,\\
\gamma^\text{trade} &=& 1,~2,~5,~10,~20,\\
\gamma^\text{hold} &=& 1,
\EEAS
a total of 45 back-test simulations.
(Logarithmic spacing is common in hyper-parameter searches.)

\begin{figure}
\begin{center}
\includegraphics[width=1.0\textwidth]{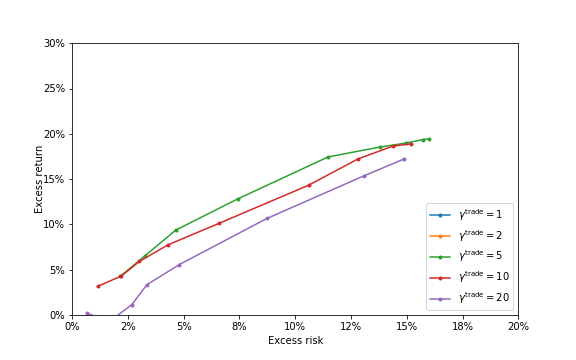}
\end{center}
\caption{SPO example, coarse hyper-parameter grid search.
(Some curves
do not fit in the plot.)
}
\label{fig:spo-frontier}
\end{figure}

Figure \ref{fig:spo-frontier} shows mean excess portfolio return $\overline{\Rep}$
versus excess volatility $\sigma^\mathrm{e}$
(defined in \S\ref{s-relativemetrics}),
for these combinations of parameters. For each
value of $\gamma^\text{trade}$, we connect with a line
the points corresponding to the different values of
$\gamma^\text{risk}$, obtaining a
risk-return trade-off curve for that choice of $\gamma^\text{trade}$
and $\gamma^\text{hold}$.
These show the expected trade-off between mean return and risk.
We see that the choice of trading aversion parameter is critical:
For some values of $\gamma^\text{trade}$ the results are so poor that
the resulting curve does not even fit in the plotting area.
Values of $\gamma^\text{trade}$ around 5
seem to give the best results.

\begin{figure}
\begin{center}
\includegraphics[width=1.0\textwidth]{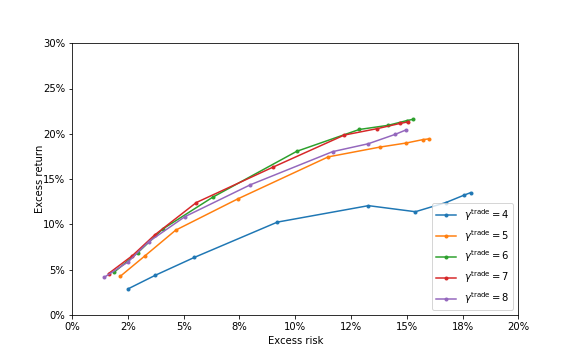}
\end{center}
\caption{SPO example, fine hyper-parameter grid search.
}
\label{fig:spo-frontier-fine}
\end{figure}

We then perform a fine hyper-parameter search,
focusing on trade aversion parameters values around
$\gamma^\text{trade} = 5$,
\[
\gamma^\text{trade} =4,~ 5,~ 6,~ 7,~ 8,
\]
and the same values of $\gamma^\mathrm{risk}$ and $\gamma^\mathrm{hold}$.
Figure \ref{fig:spo-frontier-fine} shows the resulting curves of excess return
versus excess risk. A value around $\gamma^\text{trade}=6$ seems to be best.

\begin{figure}
\begin{center}
\includegraphics[width=1.0\textwidth]{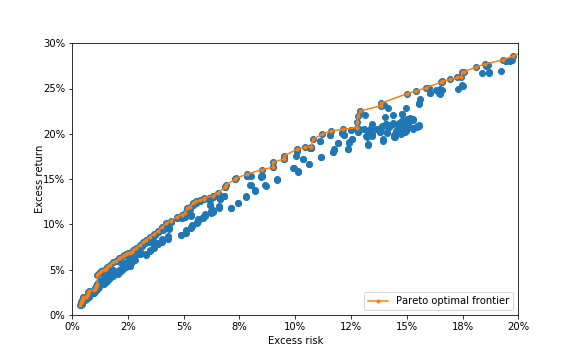}
\end{center}
\caption{SPO example, grid search over 410 hyper-parameter combinations.
The line connects the Pareto optimal points.}
\label{fig:spo-pareto}
\end{figure}

For our last set of simulations
we use a finer range of risk aversion parameters,
focus on an even narrower range of the trading aversion parameter,
and also vary the hold aversion parameter.
We test all combinations of
\BEAS
\gamma^\text{risk} &=& 0.1, ~0.178,~ 0.316, ~0.562, ~1, ~2, ~3, ~6, ~10, ~18, ~32, ~56,\\
&& \qquad ~100, ~178, ~316, ~562, ~1000,\\
\gamma^\text{trade} &=& 5.5, ~6,~6.5,~7,~7.5,~8,\\
\gamma^\text{hold} &=& 0.1, ~1, ~10,~ 100, ~ 1000,
\EEAS
a total of 410 back-test simulations.
The results are plotted in
figure \ref{fig:spo-pareto} as points in the risk-return plane.
The Pareto optimal points, \ie, those with the lowest risk
for a given level of return, are connected by a line.
Table \ref{tab:spo_pareto} lists a selection of the Pareto optimal points,
giving the associated hyper-parameter values.

\begin{table}
\begin{center}
\begin{tabular}{rrl|ll}
\toprule
&&&Excess&Excess\\
$\gamma^\mathrm{risk}$ &  $\gamma^\mathrm{trade}$ & $\gamma^\mathrm{hold}$ &  return &    risk \\
\midrule
 1000.00 &                      8.0 &                    100 &   1.33\% &   0.39\% \\
  562.00 &                      6.0 &                    100 &   2.49\% &   0.74\% \\
  316.00 &                      7.0 &                    100 &   2.98\% &   1.02\% \\
 1000.00 &                      7.5 &                     10 &   4.64\% &   1.22\% \\
  562.00 &                      8.0 &                     10 &   5.31\% &   1.56\% \\
  316.00 &                      7.5 &                     10 &   6.53\% &   2.27\% \\
  316.00 &                      6.5 &                     10 &   6.88\% &   2.61\% \\
  178.00 &                      6.5 &                     10 &   8.04\% &   3.20\% \\
  100.00 &                      8.0 &                     10 &   8.26\% &   3.32\% \\
   32.00 &                      7.0 &                     10 &  12.35\% &   5.43\% \\
   18.00 &                      6.5 &                    0.1 &  14.96\% &   7.32\% \\
    6.00 &                      7.5 &                     10 &  18.51\% &  10.44\% \\
    2.00 &                      6.5 &                     10 &  23.40\% &  13.87\% \\
    0.32 &                      6.5 &                     10 &  26.79\% &  17.50\% \\
    0.18 &                      7.0 &                     10 &  28.16\% &  19.30\% \\
\bottomrule
\end{tabular}
\end{center}
\caption{SPO example, selection of Pareto optimal points (ordered by increasing risk and return).}
\label{tab:spo_pareto}
\end{table}

From this curve and table we can make some interesting observations.
The first is that we do substantially better with large values of
the holding cost multiplier parameter compared to $\gamma^\mathrm{hold}=1$,
even though the actual hold costs (used by the simulator to update the portfolio
each day) are very small, one basis point.  This is a good example of
regularization in SPO; our large holding cost multiplier parameter
tells the SPO algorithm to avoid short positions, and the result is that
the overall portfolio performance is better.

It is hardly surprising
that the risk aversion parameter varies over this selection of Pareto
optimal points; after all,
this is the parameter most directly related to the risk-return trade-off.
One surprise is that the value of the hold aversion hyper-parameter
varies considerably as well.

In practice, we would back-test many more combinations of these three
hyper-parameters.
Indeed we would also carry out back-tests varying combinations of
other parameters in the SPO algorithm, for example the leverage,
or the individual terms in transaction cost functions.  In addition, we
would carry out stress-tests and other what-if simulations, to get an idea
of how our SPO algorithm might perform in other, or more stressful,
market conditions.
(This would be especially appropriate given our choice of
back-test date range, which was entirely a bull market.)
Since these back-tests can be carried out in parallel, there is no
reason to not carry out a large number of them.

\section{Multi-period optimization}
In this section we show the simplest possible example
of the multi-period optimization model developed in chapter \ref{s-mpo},
using planning horizon $H=2$.  This means that in each time period
the MPO algorithm plans both current day and next day trades, and then
executes only the current day trades.
As a practical matter, we would not expect a great performance improvement
over SPO using a planning horizon of $H=2$ days compared to SPO, which uses
$H=1$ day.  Our point here is to demonstrate that it is different.

The simulations are carried out using the market
data described in \S\ref{s-example-data}.
The portfolio starts with total value $v_1 =$ \$100M and uniform allocation
$w_1 = (\ones/n,0)$.  We impose a leverage constraint of $L^\mathrm{max}=3$.
The risk model is the same one used in the SPO example.  The volume and volatility
estimates (for both the current and next period)
are also the same as those used in the SPO example.

\paragraph{Return forecasts.}
We use the same return forecast we generated for the previous example,
but at every time period we provide both the forecast for the current time period
and the one for the next:
\[
\hat r_{t|t} =  \hat r_t, \quad
\hat r_{t+1|t} =  \hat r_{t+1},
\]
where $\hat r_t$ and $\hat r_{t+1}$ are the same ones used in the
SPO example, given in~(\ref{e-SPO-return-forecast}).
The MPO trading algorithm thus sees each return forecast twice,
$\hat r_{t+1} = \hat r_{t+1|t} = \hat r_{t+1|t+1}$,
\ie, today's forecast of tomorrow's return is the same as tomorrow's forecast
of tomorrow's return.

As in the SPO case, this is clearly not a practical forecast, since it uses
the realized return.
In addition, in a real setting the return forecast would be updated at
every time period,
so that $\hat r_{t+1|t} \neq \hat r_{t+1|t+1}$.
Our goal in choosing these simulated return forecasts is to have ones that are
similar to the ones used in the SPO example, in order to compare the
results of the two optimization procedures.

\paragraph{Back-tests.}
We carry out multiple back-test simulations varying the parameters
$\gamma^\text{risk}$, $\gamma^\text{trade}$, and $\gamma^\text{hold}$.
We first perform a coarse grid search in the hyper-parameter space,
with the same parameters as in the SPO example.
We test all combinations of
\BEAS
\gamma^\text{risk} &=& 0.1,~0.3,~1,~3,~10,~30,~100,~300,~1000,\\
\gamma^\text{trade} &=& 1,~2,~5,~10,~20,\\
\gamma^\text{hold} &=& 1,
\EEAS
a total of 45 back-test simulations.

\begin{figure}
\begin{center}
\includegraphics[width=1.0\textwidth]{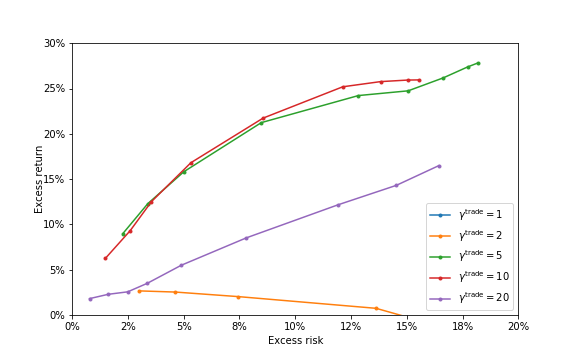}
\end{center}
\caption{MPO example, coarse hyper-parameter grid search.}
\label{fig:mpo-frontier}
\end{figure}

The results are shown in figure \ref{fig:mpo-frontier},
where we plot mean excess portfolio return $\overline{\Rep}$
versus excess risk $\sigma^\mathrm{e}$.
For some trading aversion parameter values
the results were so bad that they did not fit in the plotting area.

\begin{figure}
\begin{center}
\includegraphics[width=1.0\textwidth]{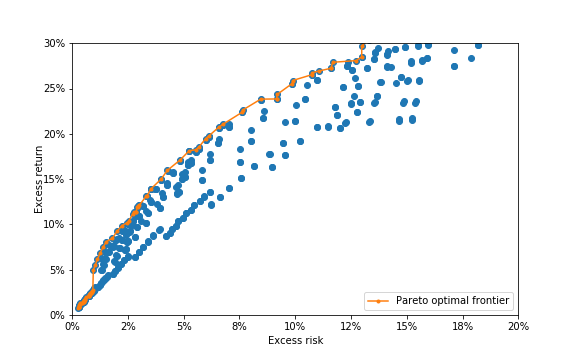}
\end{center}
\caption{MPO example, grid search over 390 hyper-parameter combinations.
The line connects the Pareto optimal points.}
\label{fig:mpo-pareto}
\end{figure}

We then perform a more accurate hyper-parameter search
using a finer range for $\gamma^\text{risk}$,
focusing on the values around $\gamma^\text{trade}=10$,
and also varying the hold aversion parameter.
We test all combinations of
\BEAS
\gamma^\text{risk} &=& 1, ~2, ~3, ~6, ~10, ~18, ~32, ~56, ~100, ~178, ~316, ~562, ~1000,\\
\gamma^\text{trade} &=& 7,~8,~9,~10,~11,~12,\\
\gamma^\text{hold} &=& 0.1,~1, ~10,~ 100, ~ 1000,
\EEAS
for a total of 390 back-test simulations.
The results are plotted in
figure \ref{fig:mpo-pareto} as points in the risk-return plane.
The Pareto optimal points are connected by a line.

\begin{figure}
\begin{center}
\includegraphics[width=1.0\textwidth]{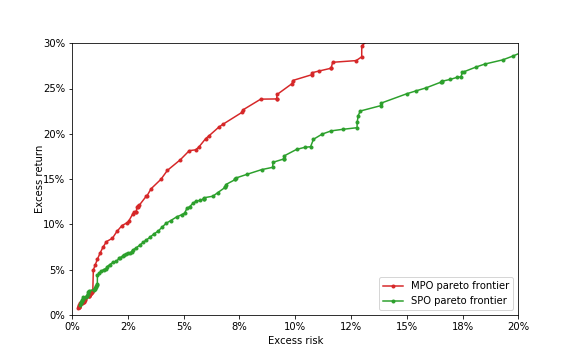}
\end{center}
\caption{Pareto optimal frontiers for SPO and MPO.}
\label{fig:spo-mpo-pareto}
\end{figure}

Finally we compare the results obtained with the SPO and MPO
examples. Figure \ref{fig:spo-mpo-pareto}
shows the Pareto optimal frontiers for both cases.
We see that the MPO method has a substantial advantage over the SPO method,
mostly explained by the advantage of a forecast for tomorrow's, as well as
today's, return.

\section{Simulation time}
Here we give some rough idea of the computation time required to
carry out the simulation examples shown above, focussing on the
SPO case.
The back-test simulation is single-threaded, so multiple back-tests
can be carried out on separate threads.

\begin{figure}
\begin{center}
\includegraphics[width=1.0\textwidth]{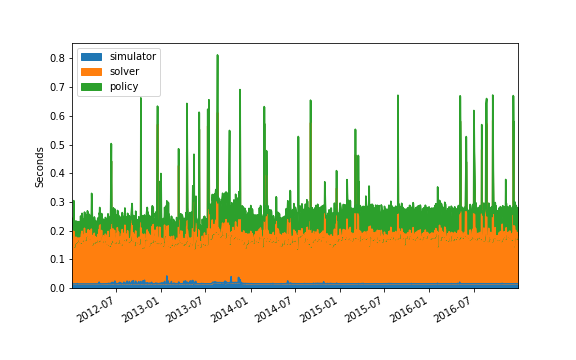}
\end{center}
\caption{Execution time for each day for one SPO back-test.}
\label{fig:execution-time}
\end{figure}

Figure \ref{fig:execution-time} gives the breakdown of execution time
for a back-test, showing the time taken for each step of simulation,
broken down into
the simulator, the numerical solver, and the rest of the
policy (data management and CVXPY manipulations).
We can see that simulating one day takes around 0.25 seconds,
so a back test over 5 years
takes around 5 minutes.
The bulk of this (around 0.15 seconds) is the optimization carried out each day.
The simulator time is, as expected, negligible.

We carried out the multiple back-tests using a 32 core machine that can execute
64 threads simultaneously.
To carry out 410 back-tests, which entails solving around a half million convex
optimization problems, thus takes around thirty minutes. (In fact, it takes a bit
longer, due to system overhead.)

We close by making a few comments about these optimization times.  First, they can
be considerably reduced by avoiding the $3/2$-power transaction cost terms, which
slow the optimizer.  By replacing these terms with square transaction cost terms,
we can obtain a speedup of more than a factor of two.
Replacing the default generic solver ECOS \cite{domahidi2013ecos} used in CVXPY
with a custom solver, such as one based on
operator-splitting methods \cite{boyd2011distributed},
would result in an even more substantial speedup.

\backmatter

\bibliography{cvx_portfolio}

\end{document}